\documentclass[useAMS,usenatbib]{mn2e}

\usepackage{graphicx}

\def\jh{\mbox{$(J-H)$}}

\def\jk{\mbox{$(J-K_s)$}}

\def\ebv{\mbox{$E(B-V)$}}

\def\ejh{\mbox{$E(J-H)$}}

\def\rc{\mbox{$R_{\rm core}$}}
\def\rl{\mbox{$R_{\rm lim}$}}

\def\ms{\mbox{$M_\odot$}}
\def\ds{\mbox{$d_\odot$}}
\def\dgc{\mbox{$d_{\rm GC}$}}

\def\jj{\mbox{$J$}}
\def\hh{\mbox{$H$}}
\def\ks{\mbox{$K_s$}}
\def\tr{\mbox{$t_{\rm rel}$}}

\title[Open clusters in dense fields]{Open clusters in dense fields: the importance of field-star 
decontamination for NGC\,5715, Lyng\aa\,4, Lyng\aa\,9, Trumpler\,23, Trumpler\,26 and Czernik\,37}

\author[C. Bonatto and E. Bica]{C. Bonatto${^1}$ and E. Bica${^1}$\\
$^1$Departamento de Astronomia, Universidade Federal do Rio Grande do Sul, Av. Bento
Gon\c{c}alves 9500\\ Porto Alegre 91501-970, RS, Brazil}

\begin{document}

\pagerange{\pageref{firstpage}--\pageref{lastpage}} 

\maketitle

\label{firstpage}

\begin{abstract}
Star clusters projected onto dense stellar fields in general require field star decontamination
to confirm their nature and derive intrinsic photometric and structural fundamental parameters.
The present work focusses on the open clusters or candidates NGC\,5715, Lyng\aa\,4, Lyng\aa\,9,
Trumpler\,23, Trumpler\,26 and Czernik\,37 which are projected within $317^\circ\la\ell\la 2.2^\circ$ 
and $|b|\la2.8^\circ$, against crowded bulge and/or disc fields. To tackle these difficult objects
we develop a CMD field star decontamination algorithm based on 2MASS \jj, \jh\ and \jk\ data,
and respective uncertainties, to detect cluster star excesses over the background. On the other
hand, colour-magnitude filters are used to build stellar radial density profiles and mass 
functions. The results convey compelling evidence that Lyng\aa\,9 and Czernik\,37 are intermediate-age 
open clusters, and their fundamental parameters are measured for the first time. Trumpler\,23 is a 
particularly challenging object, since besides high background level, its field presents variable 
absorption in near-IR bands. We confirm it to be an intermediate-age open cluster. Trumpler\,26 is 
studied in more detail than in previous works, while NGC\,5715 and Lyng\aa\,4 have fundamental parameters
determined for the first time. These open clusters are located $0.9 - 1.6$\,kpc within the Solar Circle,
with ages similar to that of the Hyades. Structurally, they are well described by King profiles. In all 
cases, core and limiting radii are significantly smaller than those of nearby open clusters outside the 
Solar circle. We test the effect of background levels on cluster radii determinations by means of 
simulations. They indicate that for central clusters, radii may be underestimated by about 10--20\%, 
which suggests that the small sizes measured for the present sample reflect as well intrinsic properties 
related to dynamical evolution effects. The objects probably have been suffering important tidal effects 
that may have accelerated dynamical evolution, especially in Czernik\,37, the innermost object.
\end{abstract}


\section{Introduction}
\label{Intro}

Colour-magnitude diagrams (CMDs) of most open clusters (OCs) feature age-dependent stellar sequences
such as fractions of the main sequence (MS), turnoff (TO) and giant branch (GB), that provide essential
information to derive their reddening, age and distance from the Sun. In this sense, OCs have been used
as probes of Galactic disc properties (\citealt{Lynga82}; \citealt{JP94}; \citealt{Friel95}; 
\citealt{DiskProp}; \citealt{Pi2006}).

However, the proximity of most OCs to the plane and the corresponding strong reddening and
field star (FS) contamination, especially for OCs projected against the central parts of the Galaxy,
usually restrict this analysis to the more populous OCs and/or those located a few kpc from the Sun
(\citealt{DiskProp}, and references therein). Using a sample of 654 OCs with available
fundamental parameters, \citet{DiskProp} found that a large fraction of the intrinsically faint and/or
distant OCs must be drowned in the field, particularly in bulge/disc directions. They estimate
a total population of $\sim10^5$ OCs in the Galaxy, in agreement with \citet{Pi2006}.

Because they are minority in catalogues, OCs projected towards the central parts of the Galaxy are of
particular interest. Studies like the present one can provide means to unambiguously unveil the nature 
of such potential candidates, which is essential to establish the fraction of star clusters as 
compared to statistical fluctuations of the dense stellar field in those directions. 

Reliable fundamental parameters of unstudied OCs are important both to disc studies and to constrain 
theories of molecular cloud fragmentation, star formation, as well as stellar and dynamical evolutions.
Structural and dynamical-related parameters of star clusters can be used to investigate whether 
the apparent scarcity of OCs inside the Solar Circle is due to observational limitations in dense stellar 
fields or enhanced tidal disruption rates because of proximity to the bulge and/or higher rates of collisions 
with molecular clouds (\citealt{DiskProp}, and references therein).

As a first step a series of faint OCs were studied using near-IR \jj, \hh\ and \ks\ photometry (\citealt{CygOB2};
\citealt{3OpticalCl}; \citealt{5LowContr}) obtained from the 2MASS\footnote{The Two Micron All Sky Survey, All 
Sky data release (\citealt{2mass1997}), available at {\em http://www.ipac.caltech.edu/2mass/releases/allsky/}} 
Point Source Catalogue (PSC). 2MASS spatial and photometric uniformity has been important to derive fundamental 
parameters and probe the nature of these objects, because it allows extraction of large surrounding fields that 
provide high star-count statistics in the near-IR.

To this purpose we have developed quantitative tools to disentangle cluster and field stars in CMDs, in
particular two different kinds of filters. Basically we apply {\em (i)} FS decontamination to uncover 
cluster evolutionary sequences from the field, which is important to derive reddening, age and distance 
from the Sun, and {\em (ii)} colour-magnitude (CM) filters, which proved to be essential for building 
intrinsic stellar radial density profiles (RDPs), as well as luminosity and mass functions. In particular, 
FS decontamination constrains more the age and distance from the Sun, especially for low-latitude OCs
(\citealt{DiskProp}). These tools were applied to OCs and embedded clusters to enhance MS and/or pre-MS 
sequences with respect to the field (\citealt{M52N3960}; \citealt{N4755}; \citealt{N6611};
\citealt{5LowContr}; \citealt{DetAnalOCs}). They were useful also in the analysis of faint and/or distant 
OCs (\citealt{CygOB2}; \citealt{3OpticalCl}; \citealt{5LowContr}; \citealt{FaintOCs}). In addition, more 
constrained structural parameters such as core (\citealt{King1966a}; \citealt{King1966b}) and limiting radii 
(\rc\ and \rl, respectively), and mass function (MF) slopes have been derived from CM-filtered photometry, 
allowing inferences on cluster dynamical state (e.g. \citealt{DetAnalOCs}). 

The present work investigates photometric and structural properties of six OCs or candidates in
dense stellar fields: NGC\,5715, Lyng\aa\,4, Lyng\aa\,9, Trumpler\,23, Trumpler\,26 and Czernik\,37. 
They were included in early catalogues of star clusters (\citealt{Alter70}; \citealt{Lynga85}, and
references therein), but only recently some of them have been analysed. From 4 to 6 catalogues
(cols.~1 and 13 of Table~\ref{tab1}) considered these objects as OCs or candidates on plate
materials. Hereafter we will adopt the acronyms Ly, Tr and Cz.

The objects are projected within $\approx45^\circ$ of the Galactic Centre. NGC\,5715, Ly\,4, Tr\,23, Ly\,9 
and Tr\,26 are located in the 4th quadrant while Cz\,37 is in the 1st quadrant. We are dealing with very 
low latitude objects, all with $|b|\la2.8^\circ$. Clearly, they are ideal targets to be analysed with 
decontamination methods.

\begin{figure*}
\begin{minipage}[b]{0.50\linewidth}
\includegraphics[width=\textwidth]{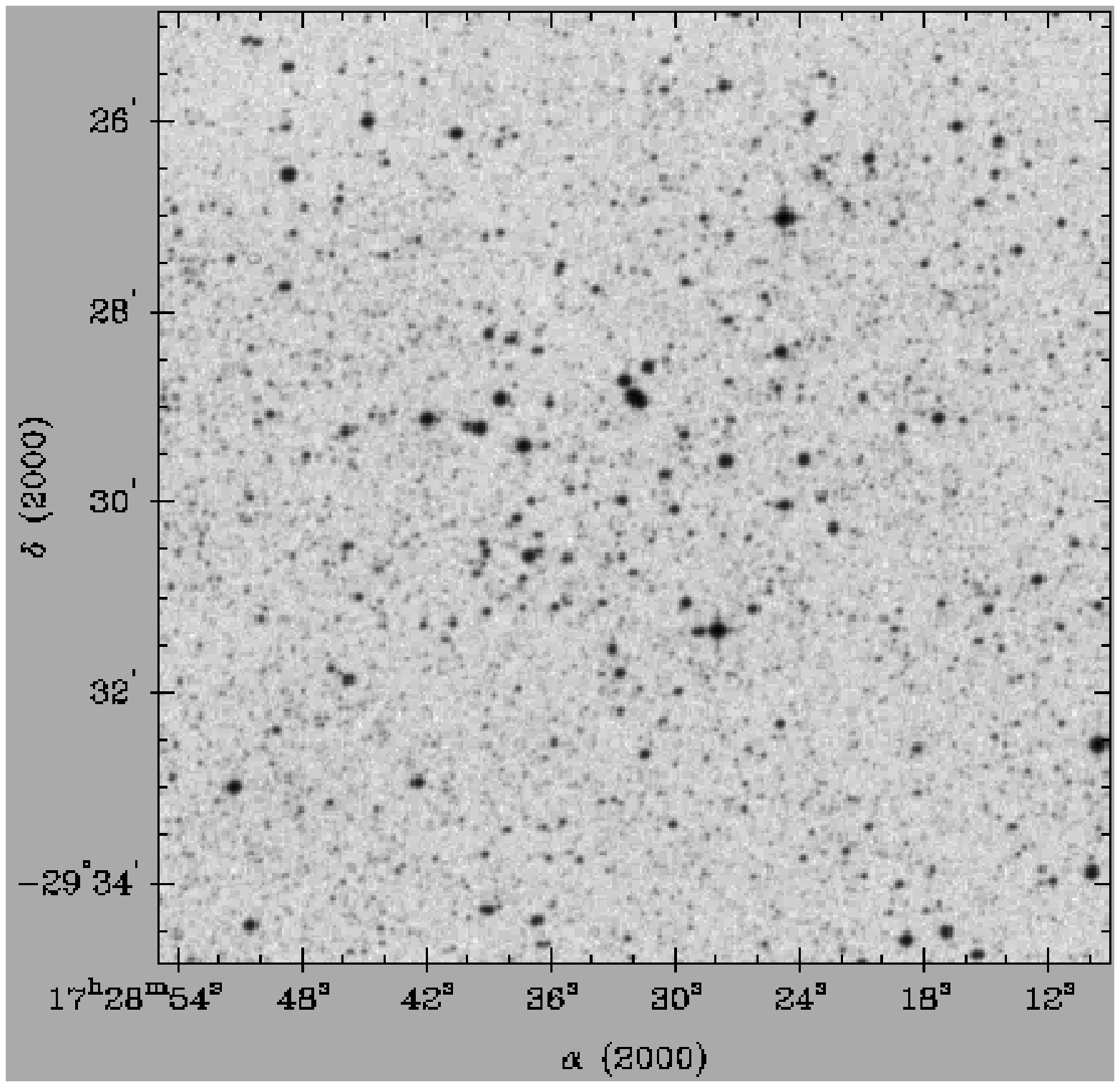}
\end{minipage}\hfill
\begin{minipage}[b]{0.50\linewidth}
\includegraphics[width=\textwidth]{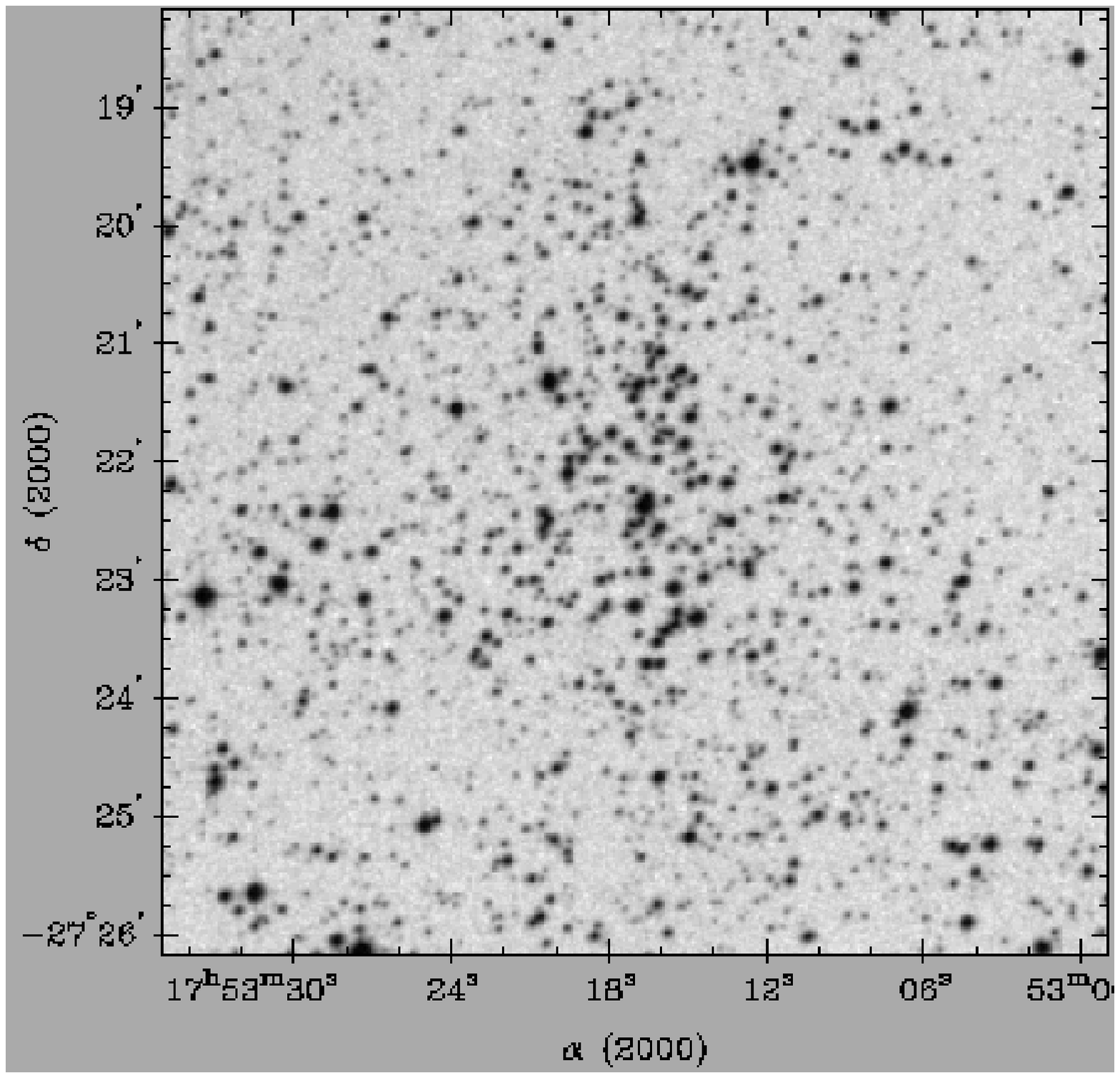}
\end{minipage}\hfill
\caption[]{Left panel: $5\arcmin\times5\arcmin$ XDSS B image of Tr\,26. Right panel: $4\arcmin\times4\arcmin$
XDSS R image of Cz\,37. Images centred on the optimized coordinates (cols.~5 and 6 of Table~\ref{tab1}).}
\label{fig1}
\end{figure*}

\begin{figure*}
\begin{minipage}[b]{0.50\linewidth}
\includegraphics[width=\textwidth]{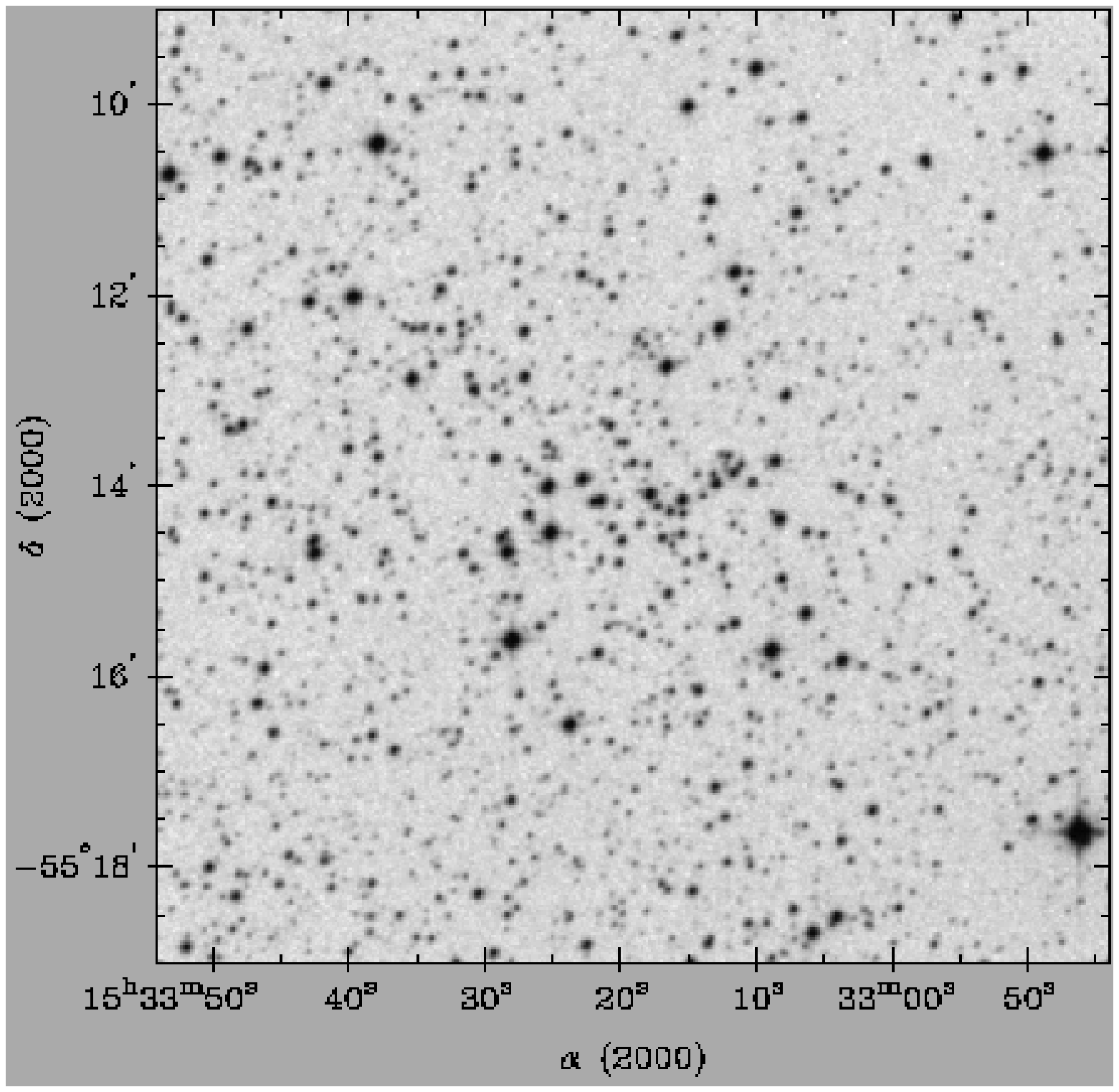}
\end{minipage}\hfill
\begin{minipage}[b]{0.50\linewidth}
\includegraphics[width=\textwidth]{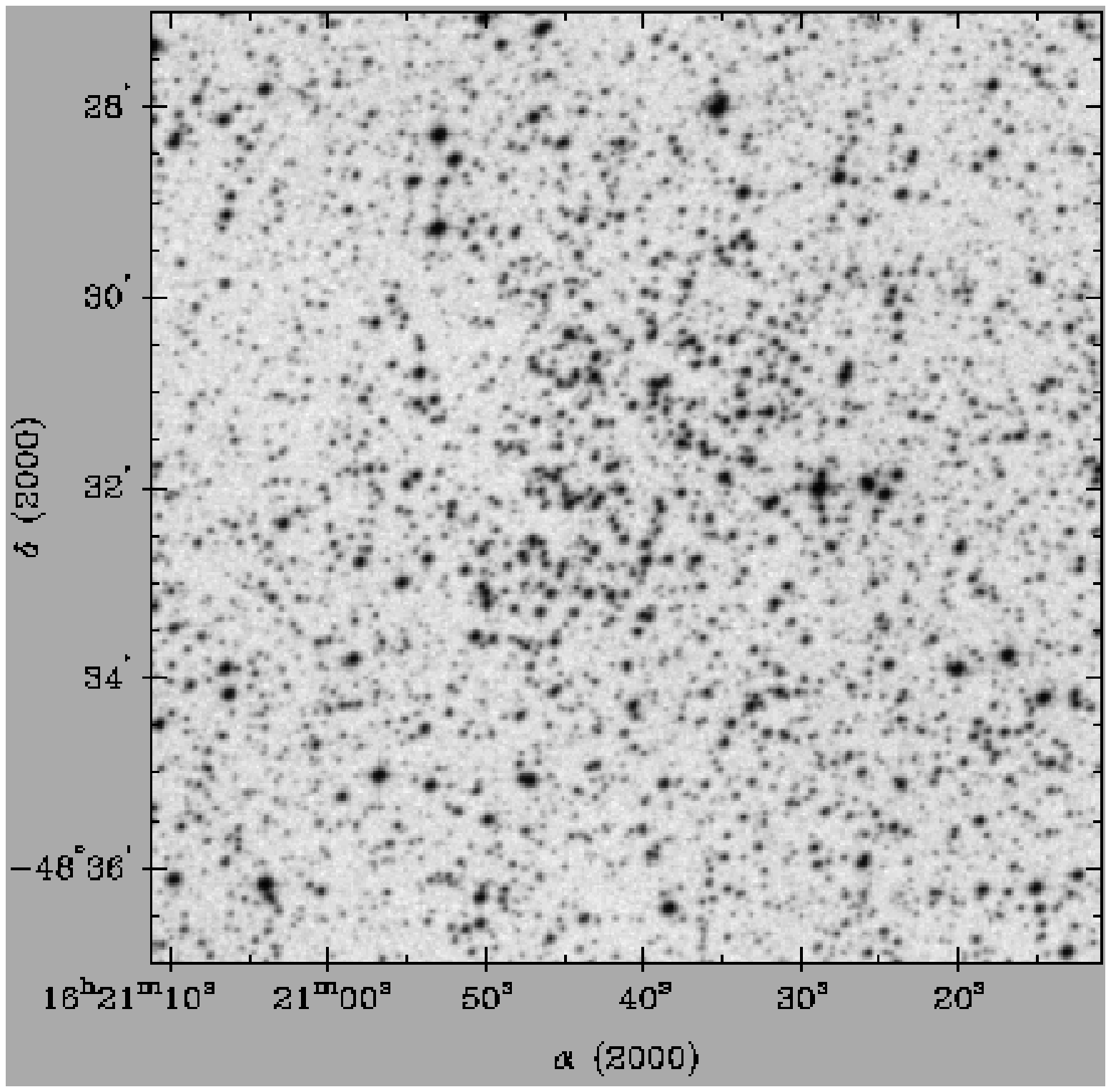}
\end{minipage}\hfill
\caption[]{Same as Fig.~\ref{fig1} for the $5\arcmin\times5\arcmin$ (XDSS R) field around
Ly\,4 (left panel) and the $5\arcmin\times5\arcmin$ (XDSS R) field around Ly\,9
(right panel).}
\label{fig2}
\end{figure*}

Tr\,23 and Tr\,26 are used to compare cluster parameters obtained by means of different decontamination 
approaches and observational data sets, and to further probe their properties. Cluster parameters for 
NGC\,5715, Ly\,4, Ly\,9 and Cz\,37 are derived for the first time.

To be considered as a high-probability star cluster, a candidate must present both a FS-decontaminated CMD 
morphology and CM-filtered stellar RDP consistent with those of typical OCs. A third criterion would be a 
cluster-like MF. However, since MFs change with cluster age and are environment dependent (\citealt{M52N3960}, 
and references therein), they do not provide enough constraints to confirm or rule out OC-candidates.

The density of stars in the direction of the present objects is unprecedented as compared to our previous
efforts to explore OCs in general. This work is expected to contribute not only with parameters 
of unstudied star clusters, but also with quantitative tools that are useful to the study of OCs 
projected in dense fields.

\begin{figure*}
\begin{minipage}[b]{0.50\linewidth}
\end{minipage}\hfill
\begin{minipage}[b]{0.50\linewidth}
\end{minipage}\hfill
\caption[]{Same as Fig.~\ref{fig1} for the $10\arcmin\times10\arcmin$ (XDSS R) field around
Tr\,23 (left panel) and the $7.5\arcmin\times7.5\arcmin$ (XDSS R) field around NGC\,5715
(right panel).}
\label{fig3}
\end{figure*}

This paper is organized as follows. Sect.~\ref{Target_OCs} contains basic properties and reviews
literature data (when available) on the objects. Sect.~\ref{2mass} discusses 2MASS data and
photometric uncertainties. In Sect.~\ref{FSD} we describe the FS decontamination algorithm to be
applied to CMDs, discuss its results and limitations, and derive fundamental cluster parameters.
Sect.~\ref{struc} presents definitions of the CM filters for each cluster, stellar RDPs and radial
 mass density profiles (MDPs), and models of the
effect of varying background levels on cluster radii determination. In Sect.~\ref{MF} MFs and cluster
mass are inferred. In Sect.~\ref{CWODS} aspects related to the structure and dynamical
state of the present objects are discussed. Concluding remarks are given in Sect.~\ref{Conclu}.

\section{The target open clusters and candidates}
\label{Target_OCs}

According to the OC catalogues WEBDA\footnote{www.univie.ac.at/webda// } and
DAML02\footnote{www.astro.iag.usp.br/~wilton/ } NGC\,5715, Ly\,9 and Cz\,37 do not have published
fundamental photometric parameters. No structural parameters are available either.

\citet{vdBH75} measured an angular diameter $D=10\arcmin$ and estimated a medium richness
for NGC\,5715. \citet{Ruprecht66} classified it as Trumpler type II\,2\,m, while \citet{Lynga82}
as III\,2\,m.

Ly\,4 was photoelectrically studied by \citet{Moffat75}, but the few observed stars did not
allow determination of a cluster sequence in the CMD. Its Trumpler type is IV\,2\,p according to
\citet{Ruprecht66} or II\,2\,m (\citealt{Lynga82}).

\citet{CJE05} photometrically studied Ly\,9, however without applying a quantitative FS decontamination. 
They concluded that Ly\,9 is an enhancement of the dense stellar field. \citet{vdBH75}
estimated an angular diameter of $D=6\arcmin$ and medium richness. Its Trumpler type is III\,1\,m
(\citealt{Ruprecht66}; \citealt{Lynga82}).

\citet{CJE05} concluded that Cz\,37 might be a real cluster superimposed on the Galactic
bulge population. \citet{vdBH75} estimated $D=3\arcmin$ and medium richness. Its
Trumpler type is III\,2\,p (\citealt{Ruprecht66}) or II\,1\,m (\citealt{Lynga82}).

\citet{CJCM06} found that Tr\,23 is an intermediate age OC within the
Solar circle that deserves further attention. They applied in the analysis the FS decontamination
method by \citet{Bertelli03}, from which they obtained $\ebv=0.83$, $\ds=2.2$\,kpc and age 1\,Gyr.
\citet{vdBH75} derived $D=6\arcmin$ and medium richness. Its Trumpler type is II\,2\,p
(\citealt{Ruprecht66}) or III\,1\,m (\citealt{Lynga82}).

\citet{Kharchenko05} employed the ASCC-2.5 catalogue with limiting magnitude $V\approx14$ to derive 
parameters for 520 open clusters, using proper motion and photometric criteria to separate probable 
members from field stars. However, owing to distance and reddening limitations, the fainter cluster
parameters rely on a few stars. For Tr\,26 they derived $\ebv=0.14$, $\ds=2.8$\,kpc, age 240\,Myr,
core and cluster radii of 4.2\arcmin\ and 9.6\arcmin, respectively. \citet{vdBH75} derived $D=6\arcmin$
and medium richness. Its Trumpler type is II\,2\,p (\citealt{Ruprecht66}) or III\,1\,m (\citealt{Lynga82}).

\begin{table*}
\caption[]{Fundamental parameters}
\label{tab1}
\tiny
\renewcommand{\tabcolsep}{0.6mm}
\renewcommand{\arraystretch}{1.5}
\begin{tabular}{lccccccccccccc}
\hline\hline

&\multicolumn{3}{c}{Measured from XDSS}&\multicolumn{8}{c}{Present results derived from 2MASS data}\\
\cline{2-4}\cline{6-13}
Cluster&$\alpha(2000)$&$\delta(2000)$&D&&$\alpha(2000)$&$\delta(2000)$&$\ell$&$b$&Age&$\ebv$&\ds&\dgc&Alternative Names\\
&(hms)&($^\circ\arcmin\arcsec$)&(\arcmin)&&(hms)&($^\circ\arcmin\arcsec$)&($^\circ$)&($^\circ$)&(Gyr)&&(kpc)&(kpc)&\\
(1)&(2)&(3)&(4)&&(5)&(6)&(7)&(8)&(9)&(10)&(11)&(12)&(13)\\
\hline
NGC\,5715&14:43:30&$-$57:34:37&6&&14:43:37.0&$-$57:34:33.6&317.54&$+$2.08&$0.8\pm0.1$&$0.42\pm0.03$&$1.5\pm0.1$
      &$6.2\pm0.1$ &Mel-128,Cr\,286,OCl-929,BH\,163,ESO\,176SC2\\

Lyng\aa\,4&15:33:19&$-$55:14:11&6&&15:33:19.0&$-$55:14:00.0&324.60&$+$0.66&$1.3\pm0.2$&$0.70\pm0.07$&$1.1\pm0.1$
      &$6.3\pm0.1$ &OCl-941,BH\,174,ESO\,147SC7 \\

Trumpler\,23&16:00:49&$-$53:32:10&5&&16:00:46.1&$-$53:31:26.4&328.85&$-$0.47&$0.9\pm0.1$&$0.58\pm0.03$&$1.9\pm0.1$
      &$5.7\pm0.1$ &Cr\,295,OCl-950,BH\,180,ESO\,178SC6\\

Lyng\aa\,9&16:20:41&$-$48:31:44&6&&16:20:41.0&$-$48:32:00.0&334.54&$+$1.07&$0.7\pm0.1$&$1.18\pm0.11$&$1.7\pm0.2$
      &$5.7\pm0.2$ &OCl-966,BH\,189,ESO\,226SC2\\

Trumpler\,26&17:28:32&$-$29:29:50&5&&17:28:32.0&$-$29:29:50.0&357.50&$+$2.84&$0.7\pm0.1$&$0.35\pm0.03$&$1.0\pm0.1$
      &$6.3\pm0.1$ &Harvard\,15,Cr\,331,OCl-1032,ESO\,454SC33\\

Czernik\,37&17:53:17&$-$27:22:10&5&&17:53:17.0&$-$27:22:10.0&2.21&$-$0.64&$0.6\pm0.1$&$1.06\pm0.03$&$1.7\pm0.1$
      &$5.6\pm0.1$ &OCl-8,BH\,253,ESO\,521SC3\\

\hline
\end{tabular}
\begin{list}{Table Notes.}
\item Cols.~2 and 3: Central coordinates measured by us on XDSS images. Col.~4: angular diameter estimated on 
XDSS images. Cols.~5-8: Optimized central coordinates (from 2MASS data). Col.~10: reddening in the object's
central region (Sect.~\ref{age}). Col.~12: \dgc\ calculated using the distance of the Sun to the Galactic centre
$R_o=7.2$\,kpc (\citealt{GCProp}).
\end{list}
\end{table*}

In Fig.~\ref{fig1} we show optical XDSS\footnote{Extracted from the Canadian Astronomy Data Centre (CADC),
at \em http://cadcwww.dao.nrc.ca/} images of Tr\,26 (left panel, B band) and Cz\,37 (right panel,
R band). Tr\,26 presents a lower contrast with respect to the background than Cz\,37. XDSS R
images of Ly\,4 (left panel) and Ly\,9 (right panel) are shown in Fig.~\ref{fig2}, while in 
Fig.~\ref{fig3} we present XDSS R images of Tr\,23 (left panel) and NGC\,5715 (right panel).
The field of Tr\,23 presents significant differential absorption, especially in the north-south
direction (Sect.~\ref{DiffRedInTr23}). All objects appear to be heavily contaminated by dense
stellar fields.

Table~\ref{tab1} provides information on the objects. Right ascension, declination and angular
diameter (Cols.~2 to 4) were visually measured by us on XDSS images (Figs.~\ref{fig1} to \ref{fig3})
as a first order approximation to the objects' centre and dimension. However, RDPs (Sect.~\ref{RDPs})
built based on the XDSS central coordinates of NGC\,5715, Tr\,23, Ly\,4 and Ly\,9 presented a dip at 
$R=0$. Consequently, new coordinates were searched to maximize the central density of stars 
by examining histograms for the number of stars in 0.5\arcmin-wide bins of right ascension and declination 
on CM-filtered photometry (Sect.~\ref{CMF}). 
The new central coordinates and the corresponding Galactic longitude and latitude are given in 
Cols.~5 to 8 of Table~\ref{tab1}. Age, central reddening, distance from the Sun and Galactocentric
distance based on 2MASS data (Sect.~\ref{age}) are given in Cols.~9 to 12. Additional designations 
in col.~13 of Table~\ref{tab1} are BH (\citealt{vdBH75}) and ESO (\citealt{Lauberts82}); previous
ones are given in \citet{Alter70}.

\section{2MASS photometry}
\label{2mass}

\jj, \hh\ and \ks\ 2MASS photometry was extracted in circular fields centred on the optimized 
coordinates of the objects (cols.~5 and 6 of Table~\ref{tab1}) using VizieR\footnote{\em
http://vizier.u-strasbg.fr/viz-bin/VizieR?-source=II/246}. Our previous experience with OC
analysis (Sect.~\ref{Intro}) shows that as long as no other populous cluster is present in the field, 
and differential absorption is not prohibitive, large extraction areas can provide the required statistics 
for a consistent FS characterization. Based on this, we used extraction radii (col.~5 of Table~\ref{tab2})
significantly larger than the respective limiting radii (Sect.~\ref{RDPs} and col.~7 of Table~\ref{tab4}) 
of the present objects. 
As photometric quality constraint, 2MASS extractions were restricted to stars with magnitudes {\em (i)} 
brighter than those of the 99.9\% Point Source Catalogue completeness limit\footnote{Following the Level\,1
Requirement, according to {\em http://www.ipac.caltech.edu/2mass/releases/allsky/doc/sec6\_5a1.html }} in 
the cluster direction, and {\em (ii)} with errors in \jj, \hh\ and \ks\ smaller than 0.2\,mag. The 99.9\%
completeness limits are different for each cluster, varying with Galactic coordinates. They are given in 
cols. 2--4 of Table~\ref{tab2}, respectively for \jj, \hh, and \ks. For reddening transformations we use
the relations $A_J/A_V=0.276$, $A_H/A_V=0.176$, $A_{K_S}/A_V=0.118$, and $A_J=2.76\times\ejh$ 
(\citealt{DSB2002}), assuming a constant total-to-selective absorption ratio $R_V=3.1$.

\begin{figure}
\resizebox{\hsize}{!}{\includegraphics[angle=0]{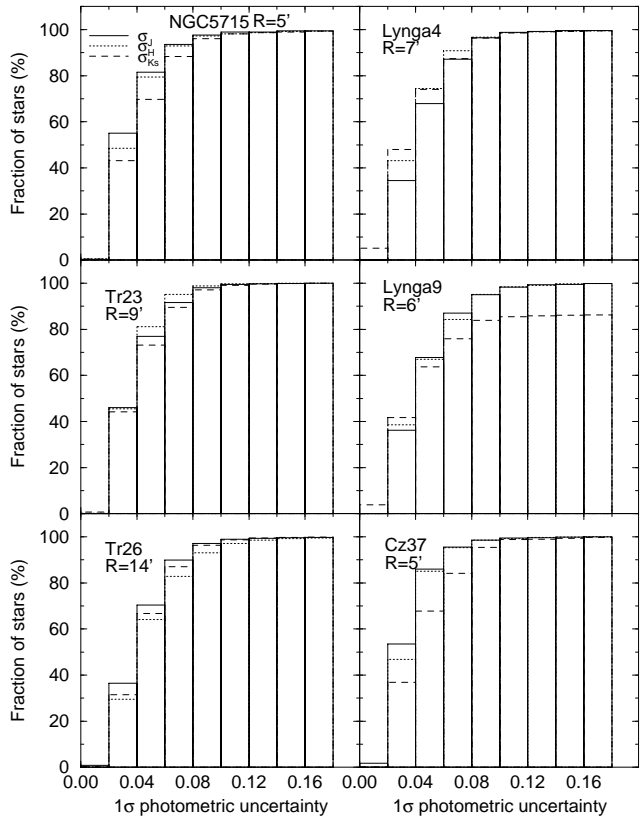}}
\caption{Quantitative evaluation of 2MASS photometric errors by means of cumulative histograms with the 
number of stars as a function of uncertainties. In all cases, most of the stars have uncertainties
smaller than 0.06\,mag.}
\label{fig4}
\end{figure}

To objectively characterize the distribution of 2MASS photometric uncertainties in the fields of the
present objects, we show Fig.~\ref{fig4} cumulative histograms with the fraction of stars as a function
of uncertainties for the 3 bands. Projected areas sampled in the histograms correspond to the respective 
limiting radius of each object (Sect.~\ref{RDPs}). The distribution of photometric uncertainties is 
similar among the fields sampled. More than $\approx80\%$ of the stars in NGC\,5715, Ly\,4, Tr\,23 and
Cz,37 have \jj\ and \hh\ uncertainties smaller than $0.06$\,mag; for \ks\ this fraction is slightly
smaller. In Tr\,26 and Ly\,9, the $0.06$\,mag fractions are $\approx70\%$. 2MASS uncertainties are 
in general larger than those of CCD photometry. However, for most stars in the present 
objects they are still small enough to provide reliable values of magnitude and colours. Besides,
for this kind of study large surrounding fields are necessary for statistical representativity 
of FSs (Sect.~\ref{FSD}) and long-base RDPs (Sect.~\ref{RDPs}), which in general are not available
in the optical.

\begin{table}
\caption[]{Details on 2MASS photometry}
\label{tab2}
\renewcommand{\tabcolsep}{1.2mm}
\renewcommand{\arraystretch}{1.2}
\begin{tabular}{lcccccc}
\hline\hline
&\multicolumn{3}{c}{Completeness limit}\\
\cline{2-4}
Cluster&\jj  &\hh  &\ks  &&$R_{ext}$&Comparison field\\
       &(mag)&(mag)&(mag)&&(\arcmin)&(\arcmin)\\
(1)&(2)&(3)&(4)&&(5)&(6)\\
\hline
NGC\,5715&15.8&15.1&14.3&&30&10--30\\
Lyng\aa\,4&15.5&15.1&14.3&&50&20--50 \\
Trumpler\,23&15.5&15.0&14.2&&40&$^\dag$\\
Lyng\aa\,9&15.5&14.8&14.0&&30&10--30\\
Trumpler\,26&14.2&13.5&12.5&&50&20--50\\
Czernik\,37&15.0&14.0&13.2&&40&10--40\\
\hline
\end{tabular}
\begin{list}{Table Notes.}
\item Column~5: 2MASS extraction radius. $(^\dag)$ - Because of heavy differential absorption,
the comparison field of Tr\,23 was taken from two $R=5\arcmin$ circular regions located at 
20\arcmin\ east and west of the cluster centre (Sect.~\ref{DiffRedInTr23}).
\end{list}
\end{table}

\begin{figure}
\resizebox{\hsize}{!}{\includegraphics[angle=0]{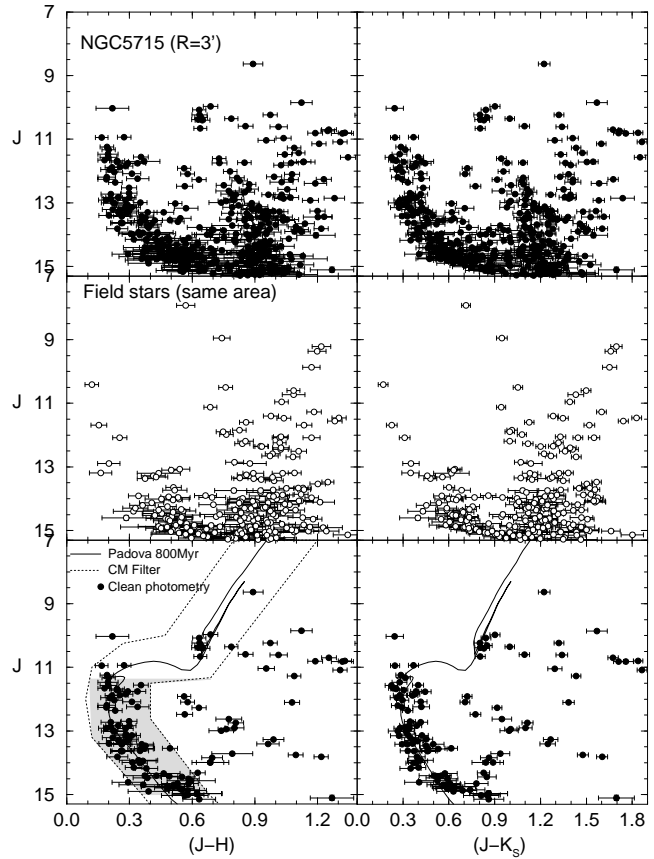}}
\caption{Top-left panel: observed $\jj\times\jh$ CMD of a central ($R=3\arcmin$) region of
NGC\,5715, which shows some contamination of bulge stars, especially for $\jh\ga 0.75$.
Middle-left panel: FSs taken from the comparison field with a projected area equal to that in the 
above panel. Bottom-left panel: FS decontaminated CMD together with the fit of the 800\,Myr 
Padova isochrone (solid line) and the CM filter (Sect.~\ref{CMF}) used to isolate cluster MS/evolved 
stars (dotted line). Completeness limit for this cluster ($\jj=15.8$) is just below the bottom of the 
CMDs. The magnitude range for MF purposes is shown as a shaded region about the MS. Right panels: the 
same for the $\jj\times\jk$ CMD. The giant clump and a significant fraction of the MS are conspicuous
in both colours.}
\label{fig5}
\end{figure}

In Fig.~\ref{fig5} we present the analysis of NGC\,5715, by means of $\jj\times\jh$ and $\jj\times\jk$
CMDs extracted from a central ($R=3\arcmin$) region. The extension of this extraction corresponds to
about twice the core radius and somewhat larger than half the limiting radius (Sect.~\ref{struc}). This
extraction provides an adequate contrast (in terms of density of stars) between CMD sequences of the
object and offset field. Photometry was limited in magnitude according to
cols.~2 -- 4 of Table~\ref{tab2}; error bars show that photometric uncertainties, although increasing
for faint stars (Fig.~\ref{fig4}), are not large to the point of producing smeared CMDs. Bulge stars 
are conspicuous in the field of NGC\,5715, as shown in both CMDs (top panels), especially for $\jh\ga0.75$ 
and $\jk\ga1.0$. However, comparison of the central CMDs (top panels) with those of the equal area offset 
field (middle panels) suggests an excess of stars for bluer colours, which is indicative of a MS. Evidence
of a giant clump is also present in both CMDs. The statistical significance of this excess is further 
detached on the FS-decontaminated CMD morphology (bottom panels of Fig.~\ref{fig5}). We explore 
FS-decontaminated CMD morphology further in Sect.~\ref{FSD}.

\begin{figure}
\resizebox{\hsize}{!}{\includegraphics[angle=0]{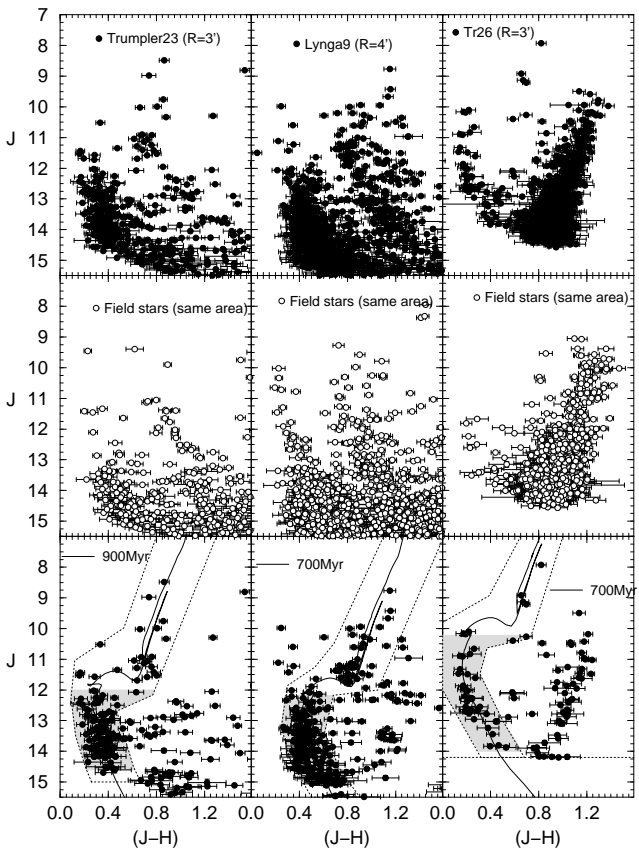}}
\caption{Same as Fig.~\ref{fig5} for the $\jj\times\jh$ CMDs of Tr\,23 (left panels), Ly\,9 (middle
panels), and Tr\,26 (right panels). Completeness limits are $\jj=15.5$ for Tr\,23 and Ly\,9,
and $\jj=14.2$ for Tr\,26. Isochrone ages are 900\,Myr (Tr\,23) and 700\,Myr (Tr\,26 and Ly\,9).}
\label{fig6}
\end{figure}

\begin{figure}
\resizebox{\hsize}{!}{\includegraphics[angle=0]{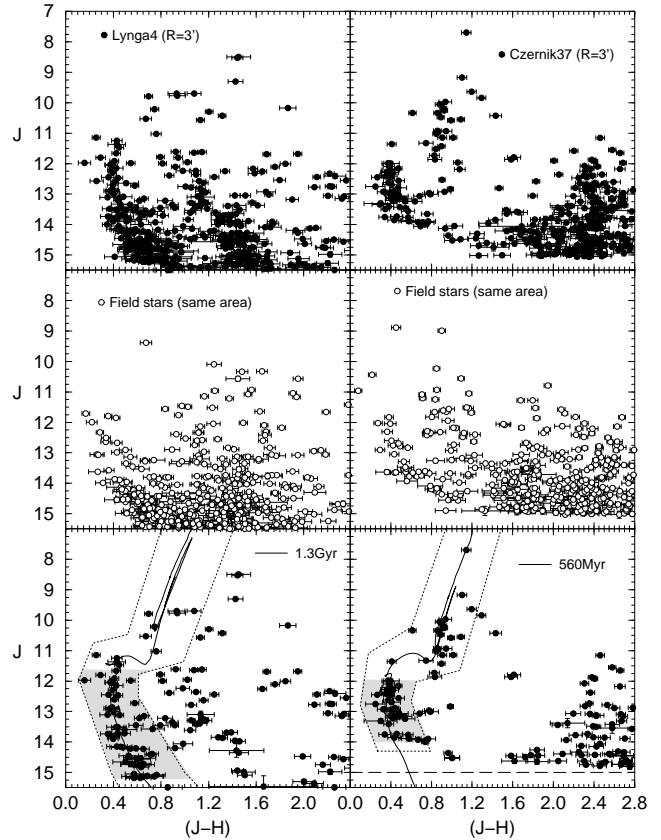}}
\caption{Same as Fig.~\ref{fig5} for the $\jj\times\jh$ CMDs of Ly\,4 (left panels) and Cz\,37 (right
panels). Completeness limits are $\jj=15.5$ for Ly\,4 and $\jj=15.0$ for Cz\,37. Isochrone ages are
1.3\,Gyr (Ly\,4) and 560\,Myr (Cz\,37).}
\label{fig7}
\end{figure}

Similar analyses involving $\jj\times\jh$ and $\jj\times\jk$ CMDs were applied to the remaining 
objects. In all cases we show CMDs of extractions taken from regions with radius intermediate between 
core and limiting radii. For the sake of simplicity only the $\jj\times\jh$ CMDs are shown in 
Figs.~\ref{fig6} and \ref{fig7}. In all cases, varying proportions of bulge stars can be seen in the 
observed CMDs (top panels). Disentangling cluster and field sequences in such dense fields requires a 
quantitative method (Sect.~\ref{FSD}).

\subsection{Spatially variable absorption in the field of Tr\,23}
\label{DiffRedInTr23}

Some peculiarities associated with variable absorption that affects star sequences in the 
CMD of Tr\,23 are important to be discussed.

Fig.~\ref{fig3} indicates the presence of important variable absorption in the optical
field of Tr\,23. To investigate its spatial distribution we plot in the top panels of
Fig.~\ref{fig8} the 2MASS position of each star (with respect to the central coordinates -
Table~\ref{tab1}) both in the observed (left) and CM-filtered (Sect.~\ref{CMF}) photometry
(right). The distribution of stars in both panels are affected by differential absorption,
especially in the north-south direction.

\begin{figure}
\caption{Top-left panel: projected position of the stars in the field of Tr\,23. Top-right:
same as before for the CM-filtered photometry; core and limiting radii, as well as the east/west
comparison fields are marked off. Bottom-left: north-south profile extracted within the central
stripe located at $-5\le\alpha(\arcmin)\le+5$. Bottom-right: east-west profile extracted within 
the stripe $-5\le\delta(\arcmin)\le+5$.}
\label{fig8}
\end{figure}

This trend is confirmed by the linear profiles (number of stars per bin) extracted along declination 
(bottom-left panel) and right-ascension (bottom-right) directions, centred on the coordinates of Tr\,23
(Table~\ref{tab1}). The field along the east-west direction is more uniform than the north-south one. To 
minimize differential absorption effects, we take as comparison field the average of 2 circular regions 
of 5\arcmin\ in radius shifted 20\arcmin\ east and west off the centre of Tr\,23. CMDs of a central 
($R=3\arcmin$) and offset field regions, as well as the FS decontaminated one of Tr\,23 are shown in 
Fig.~\ref{fig6} (left panels).

\section{Field-star decontamination in CMDs and cluster fundamental parameters}
\label{FSD}

Important field star contamination is conspicuous in the observed CMDs of the present, low-latitude objects, 
projected against bulge directions (upper panels of Figs.~\ref{fig5} -- \ref{fig7}). Quantitative FS
decontamination becomes fundamental to determine the nature of these objects, as real star 
clusters or random fluctuations of the dense background stellar density, i.e. asterisms.

Disentangling cluster and field stars is an important - but difficult - task, especially if dense
fields and poorly-populated clusters are involved. Ideally, cluster member identification is more
constrained when photometric and proper motion data are available for a large number of individual stars
(e.g. \citealt{DiasCat}). However, for most star clusters, distant ones in particular, proper motion data 
are not available. Thus the task has to be done using only photometry, by means of statistical comparison 
of star samples taken from the cluster region and offset field.

\subsection{Description of the FS decontamination algorithm}
\label{Descr_FSD}

To uncover the intrinsic CMD morphology we use an upgraded version of the FS decontamination
algorithm previously applied in the analysis of low-contrast (\citealt{5LowContr}), embedded
(\citealt{N6611}), young (\citealt{N4755}), and faint (\citealt{FaintOCs}) OCs. Sampling small 
CMD regions, the algorithm 
works on a statistical basis that takes into account the relative number-densities of stars in 
a cluster region and comparison field. It can be applied to the full cluster region ($0\leq R\leq\rl$)
or inner regions such as core and/or halo. This upgraded version of the algorithm works with 3 
dimensions, the \jj\ magnitude and the \jh\ and \jk\ colours, considering as well the respective 
$1\sigma$ uncertainties in the 2MASS bands, $\sigma_J$, $\sigma_H$ and $\sigma_{K_s}$. These are
the 2MASS colours that provide the maximum variance among cluster CMD sequences for OCs of
different ages (e.g. \citealt{TheoretIsoc}).

Basically, the algorithm {\em (i)} divides the full range of magnitude and colours of a given CMD 
into a 3D grid whose cubic cells have axes along the \jj, \jh\ and \jk\ directions, 
{\em (ii)} computes the expected number-density of FSs in each cell based on the number of comparison 
field stars with magnitude and colours compatible with those of the cell, and {\em (iii)} randomly 
subtracts the expected number of FSs from each cell. Consequently, this method is sensitive to local 
variations of FS contamination with magnitude and colours. To simplify notation we use the definitions
$\chi=\jj$, $\xi=\jh$ and $\zeta=\jk$.

Cell dimensions are $\Delta\chi=0.5$ and $\Delta\xi=\Delta\zeta=0.2$, except in Cz\,37 where
$\Delta\xi=\Delta\zeta=0.15$. These values are large enough to allow sufficient star-count statistics in
individual cells and small enough to preserve the morphology of different CMD evolutionary sequences. 
Besides, 2MASS photometric uncertainties for most stars of the present objects are considerably smaller
than the adopted cell dimensions (Sect.~\ref{2mass} and Fig.~\ref{fig4}).

To illustrate the process, consider a CMD cell whose sides in the $(\chi,\xi,\zeta)$ space have coordinates
$(\chi_c\pm\Delta\chi/2,\xi_c\pm\Delta\xi/2,\zeta_c\pm\Delta\zeta/2)$, where $(\chi_c,\xi_c,\zeta_c)$
are the cell's central coordinates. We use Gaussian distributions of magnitude\footnote{e.g.
$P(\jj,\bar J)=\frac{1}{\sqrt{2\pi}\sigma_J}\,e^{{-1/2}\left(\frac{J-\bar J}{\sigma_J}\right)^2}$}
and colours to compute the probability of a star with CMD coordinates
$(\bar\chi\pm\sigma_\chi,\bar\xi\pm\sigma_\xi,\bar\zeta\pm\sigma_\zeta)$ to be found within that
cell. In this way, computations take into account magnitude and colour uncertainties, in the sense that
stars with large uncertainties may have a non-negligible probability of populating more than one CMD cell. 
We assume that probable cluster stars are distributed in the region $0<R<\rl$, with projected area $A_{cl}$, 
while FSs are taken from the offset field region ($R>\rl$), with projected area $A_{fs}$. The expected FS
number-density ($\rho_{fs}^{cell}$) in a given cell is computed by summing up the individual probability
($P_{fs}^{cell}$) of all offset field stars ($N_{fs}$) to be found in the cell, divided by the offset
field area, $\rho_{fs}^{cell}=P_{fs}^{cell}/A_{fs}$, where

\[ P_{fs}^{cell}=\sum_{i=1}^{N_{fs}}\int\int\int P_i(\chi,\bar\chi_i;\xi,\bar\xi_i;\zeta,\bar\zeta_i)
\,d\chi\,d\xi\,d\zeta~~~~.\]

$P_i(\chi,\bar\chi_i;\xi,\bar\xi_i;\zeta,\bar\zeta_i)$ represents the probability of the $i$-th offset 
field star, with CMD coordinates $(\bar\chi_i,\bar\xi_i,\bar\zeta_i)$ and uncertainties $(\sigma_{\chi_i},
\sigma_{\xi_i},\sigma_{\zeta_i}$), to have the magnitude and colours $(\chi, \xi,\zeta)$.
Integrals are carried over the cell's extension in each dimension,
$\chi_c-\Delta\chi/2\leq\chi\leq\chi_c+\Delta\chi/2$, $\xi_c-\Delta\xi/2\leq\xi\leq\xi_c+\Delta\xi/2$
and $\zeta_c-\Delta\zeta/2\leq\zeta\leq\zeta_c+\Delta\zeta/2$, respectively; they basically reduce
to error functions computed at the cell borders. 

We do the same to compute the number-density of observed (cluster $+$ FS) stars in the cell,
$\rho_{obs}^{cell}=P_{obs}^{cell}/A_{cl}$. In this case we consider only stars that are in the
cluster region ($0\leq R\leq\rl$). Thus, the expected number of FSs in the cell is given by
$n_{fs}^{cell}=\frac{\rho_{fs}^{cell}}{\rho_{obs}^{cell}}\times n_{obs}^{cell}$, where $n_{obs}^{cell}$
is the number of observed stars (at $R\leq\rl$) located in the cell. The number of probable member stars
in the cell is $n_{cl}^{cell}=n_{obs}^{cell}-n_{fs}^{cell}$. Finally, the total number of probable cluster
members is obtained by summing $n_{cl}^{cell}$ over all CMD cells, $N_{cl}= \sum_{cell}n_{cl}^{cell}$.

As a last step, to minimize potential artificialities introduced by the choice of parameters, we apply the
decontamination algorithm for 3 different grid specifications in each dimension. For instance, for a
CMD grid beginning at magnitude $J_o$ (and cell width $\Delta\jj$), we also include additional runs for
$J_o\pm\frac{1}{3}\Delta\jj$. Considering as well similar variations for the 2 colours, 
27 different outputs are obtained, from which we compute the average number of probable cluster stars
$\langle N_{cl}\rangle$. Typical standard deviations of $\langle N_{cl}\rangle$ are at the $\approx2.5\%$
level. The final FS-decontaminated CMD contains the $\langle N_{cl}\rangle$ stars with the highest
number-frequencies.

\subsection{FS subtraction efficiency}
\label{fsub}

Working with densities inevitably results in fractional numbers of FSs in some cells. Fractions, in those
cases, are rounded off to the nearest integer, however, limited to the number of observed stars present 
in the cell ($n_{fs}^{cell}\leq n_{obs}^{cell}$). To take this effect into account, we compute for each
cell the difference between the expected number of FSs and the actual number of subtracted stars. Summing 
this difference over all cells gives us an estimate of the total number of un-subtracted FSs (\~N$_{fs}$), 
as well as a measure of the global subtraction efficiency ($f_{sub}$) of the process. The un-subtracted stars 
cannot be deleted from CMDs, since \~N$_{fs}$ results from the sum of star fractions over all cells. However, 
$f_{sub}$ can be used to compute the corrected fraction of cluster member stars. Subtraction efficiency and 
corrected fraction of cluster members for the present objects are given in Table~\ref{tab3}.

Small values of $\Delta\jj$, $\Delta\jh$\ and $\Delta\jk$\ increase the frequency of cells with fractional 
values of FSs, which in turn produce small subtraction efficiencies. Large values, on the other hand, increase 
subtraction efficiencies, but tend to degrade CMD resolution, which may be necessary to disentangle cluster 
and field star sequences. The adopted values of cell dimensions (see above) are a compromise between subtraction 
efficiency and CMD resolution at the 2MASS photometric uncertainties.

In principle, inclusion of additional dimensions such as the \hh\ or \ks\ magnitudes, could produce
more constrained results. On the other hand, this would as well reduce the number of stars in the 
multi-dimensional cells and, consequently, the subtraction efficiency would become significantly 
reduced. 

\subsection{Application to the objects and results}
\label{Applic}

The output of the FS decontamination algorithm can be examined from two different, but complementary,
perspectives, i.e. {\em (i)} the dependence of FS contamination (or alternatively, fraction of probable 
cluster stars) on apparent magnitude, and {\em (ii)} distance to the cluster center.

\begin{figure}
\resizebox{\hsize}{!}{\includegraphics[angle=0]{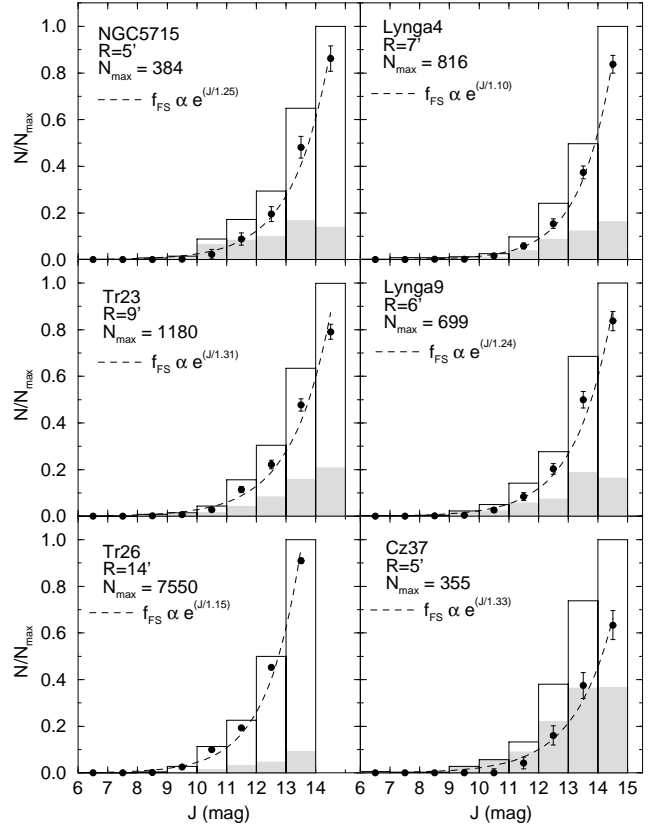}}
\caption{Fraction of stars before (empty histogram) and after (gray) FS decontamination as a
function of magnitude. Regions sampled by the histograms correspond to the limiting radius of
each cluster. For clarity, histograms were normalized to the number of stars in the highest bin.
The residual FS contribution (filled circles) is well represented by an exponential growth curve
$f_{\rm FS}(\jj)\propto\,e^{(J/\epsilon_J)}$ (dashed line).}
\label{fig9}
\end{figure}

Fig.~\ref{fig9} provides histograms that show for each object the fraction of stars before and after FS
decontamination as a 
function of \jj\ magnitude, sampled in bins of $\Delta\jj=1$\,mag. In all cases, the area sampled by the 
histograms ranges from the centre to the limiting radius (Sect.~\ref{struc}). As expected from their almost 
central directions, FS contamination in the fields of the present 6 objects dominates the observed photometry
($\ga90\%$ - Table~\ref{tab3}), increasing exponentially for faint magnitudes. Fig.~\ref{fig9} shows that 
in all cases the increase with magnitude of the residual FS contribution (subtraction of 
the probable cluster stars from the observed ones) is well represented by the exponential growth curve 
$f_{\rm FS}(\jj)\propto\,e^{(J/\epsilon_J)}$, with a magnitude-scale factor $\epsilon_J\approx1.2\pm0.1$.
Consequently, at the 2MASS completeness limits (Sect.~\ref{2mass}), FSs should represent more than 
$\approx90\%$ of the stars populating CMDs of bulge-projected open clusters. The above discussion further 
stresses the difficulties associated with the study of OCs projected against central parts of the Galaxy, 
especially to access the low MS of poorly-populated objects.

With respect to radial distribution, Table~\ref{tab3} provides statistics on the fraction of probable
cluster members (considering the full magnitude range), both for the observed and CM-filtered photometries, 
for annular extractions located at different distances from the central positions. Probable cluster member
fractions have been corrected for the respective subtraction efficiencies. In all cases, field stars dominate 
over cluster members in the observed photometry (top half of Table~\ref{tab3}), since the former contribute 
with more than $\approx90\%$ to the stars in the overall ($R\leq\rl$) CMDs of the present objects. However, 
as expected from star clusters, the fraction of probable members systematically increases for smaller radii,
reaching a maximum at the centre. In the present objects, central cluster member fractions (observed
photometry) range from $\approx11\%$ (Tr\,26) to $\approx46\%$ (Cz\,37).

The above features are enhanced in the output of the FS decontamination when applied to CM-filtered data
(bottom half of Table~\ref{tab3}). As suggested by previous works (e.g. \citealt{DetAnalOCs}; \citealt{N4755}), 
CM filters allow detection of cluster structures at larger distances from the cluster center than with 
observed photometry. The elimination of most of the obvious non-cluster stars provided by the CM filter makes 
the FS contamination drop to $\sim55-87\%$ for the present objects. This, in turn, allows a better 
determination of cluster structural parameters and dimension. Indeed, central cluster member fractions 
become significantly larger, as compared to those obtained with observed photometry, ranging from 
$\approx45\%$ (Ly\,4) to $\approx79\%$ (Tr\,26).

\begin{table*}
\caption[]{Spatial properties of the field star decontamination}
\label{tab3}
\tiny
\renewcommand{\tabcolsep}{0.75mm}
\renewcommand{\arraystretch}{1.25}
\begin{tabular}{crcccrcccrcccrcccrcccrccc}
\hline\hline
&\multicolumn{23}{c}{Observed photometry}\\
\cline{2-24}
&\multicolumn{3}{c}{NGC\,5715}&&\multicolumn{3}{c}{Ly\,4}&&\multicolumn{3}{c}{Tr\,23}&&\multicolumn{3}{c}{Ly\,9}
&&\multicolumn{3}{c}{Tr\,26}&&\multicolumn{3}{c}{Cz\,37}\\
\cline{2-4}\cline{6-8}\cline{10-12}\cline{14-16}\cline{18-20}\cline{22-24}
$\Delta R$&N&$f_{sub}$&$f_{cl}$ &&N&$f_{sub}$&$f_{cl}$ &&N&$f_{sub}$&$f_{cl}$&
&N&$f_{sub}$&$f_{cl}$&&N&$f_{sub}$&$f_{cl}$ &&N&$f_{sub}$&$f_{cl}$\\
(\arcmin)&(*)&(\%)&(\%)&&(*)&(\%)&(\%)&&(*)&(\%)&(\%)&&(*)&(\%)&(\%)&&(*)&(\%)&(\%)&&(*)&(\%)&(\%)\\
   (1)   &(2)    &(3) &(4) &&(5)  &(6)&(7) &&(8) & (9)&(10)&&(11)&(12)&(13) &&(14)&(15)&(16) &&(17)&(18)&(19)\\
\hline
0-2  &193&$84\pm3$&$26.8\pm5.8$&&218&$91\pm1$&$28.6\pm5.1$&&211&$81\pm2$&$42.4\pm4.6$&&311&$80\pm2$&$35.7\pm4.0$&&
      329&$91\pm2$&$10.8\pm5.5$&&204&$72\pm4$&$46.3\pm5.6$\\

2-4  &486&$91\pm1$&$9.0\pm4.2$&&542&$90\pm1$&$9.3\pm4.0$&&564& $90\pm2$&$28.5\pm3.5$&&722&$86\pm1$ &$10.0\pm3.5$&&
      1028&$96\pm1$ &$13.0\pm2.9$&&495&$71\pm3$ &$12.2\pm7.2$\\

4-6  &736&$89\pm1$&$0.0\pm3.8$&&892&$90\pm1$ &$6.9\pm3.1$&&843&$86\pm1$ &$18.1\pm3.3$&&1063&$85\pm1$ &$0.0\pm3.3$&&
      1656&$95\pm1$ &$8.4\pm2.8$&&851&$67\pm2$ &$8.4\pm6.4$\\

6-8  &---& &---&&1274&$91\pm1$ &$7.8\pm2.6$&&1016&$82\pm1$ &$2.8\pm3.3$&&---& &---&&
      2192&$87\pm1$ &$2.8\pm2.9$&&1203&$69\pm1$ &$0.2\pm7.4$\\

8-10 &---& &---&&1546&$90\pm1$ &$1.6\pm2.5$&&1256&$84\pm1$ &$0.0\pm3.0$&&---& &---&&
      2822&$92\pm1$ &$1.9\pm2.2$&&---& &---\\

10-12&---& &---&&---& &---&&---& &---&&---& &---&&
      3443&$91\pm1$ &$2.5\pm2.1$&&---& &---\\

12-14&---& &---&&---& &---&&---& &---&&---& &---&&
      4071&$92\pm1$ &$2.0\pm1.9$&&---& &---\\
\hline
Total &995&$90\pm1$ &$6.8\pm3.2$&&2241&$91\pm1$ &7.4$\pm2.0$&&3237&$85\pm1$ &$9.2\pm1.8$&&2096&$86\pm1$ &$7.4\pm2.2$&&
      15541&$92\pm1$ &$6.3\pm1.5$&&1149&$65\pm1$ &$7.1\pm6.9$\\
\hline
&\multicolumn{23}{c}{CM-filtered photometry}\\
\cline{2-24}
0-2  &84&$95\pm4$ &$56.3\pm5.4$&&95&$95\pm3$ &$44.9\pm6.0$&&105&$94\pm4$ &$63.0\pm4.4$&&187&$92\pm4$ &$58.6\pm3.6$&&
      36&$84\pm11$ &$79.0\pm5.2$&&50&$74\pm11$ &$78.7\pm5.7$\\

2-4  &175&$95\pm1$ &$33.9\pm5.1$&&196&$94\pm1$ &$16.9\pm6.0$&&215&$98\pm1$ &$45.6\pm4.0$&&309&$97\pm1$ &$23.0\pm4.6$&&
       45&$99\pm5$ &$42.4\pm10.5$&&55&$89\pm6$ &$36.5\pm9.7$\\

4-6  &271&$90\pm1$ &$1.0\pm7.5$&&300&$94\pm1$ &$5.1\pm5.6$&&316&$98\pm1$ &$37.0\pm3.8$&&426&$95\pm1$ &$6.1\pm4.7$&&
       60&$90\pm3$ &$24.2\pm5.5$&&85&$92\pm4$ &$31.4\pm8.6$\\

6-8  &---& &---&&445&$96\pm1$ &$11.9\pm4.3$&&329&$95\pm2$ &$16.6\pm5.4$&&620&$97\pm1$ &$10.5\pm3.8$&&
       85&$99\pm2$ &$26.6\pm8.2$&&101&$81\pm3$ &$4.8\pm10.4$\\

8-10 &---& &---&&512&$92\pm1$ &$1.4\pm4.4$&&375&$93\pm1$ &$6.4\pm5.0$&&---& &---&&
       79&$88\pm2$ &$0.0\pm12.6$&&---& &---\\

10-12&---& &---&&---& &---&&---& &---&&---& &---&&
      120&$98\pm2$ &$17.7\pm8.0$&&---& &---\\

12-14&---& &---&&---& &---&&---& &---&&---& &---&&
      111&$92\pm2$ &$0.0\pm5.4$&&---& &---\\
\hline
TF&348&$96\pm2$ &$30.8\pm4.0$&&809&$97\pm1$ &$14.8\pm3.8$&&1140&$97\pm1$ &$28.0\pm2.3$&&
      922&$97\pm1$ &$21.9\pm2.7$&&536&$98\pm1$ &$13.1\pm4.2$&&158&$92\pm3$ &$45.2\pm5.8$\\
\hline

\end{tabular}
\begin{list}{Table Notes.}
\item Cols.~2, 5, 8, 11, 14, and 17: number of stars in the region specified in col.~1. Cols.~3, 6, 9, 
12, 15, and 18: subtraction efficiency; Cols.~4, 7, 10, 13, 16 and 19: corrected fraction of probable 
member stars. Total fields (TF) correspond to extractions with radii $R=5.0\arcmin$ (NGC\,5715), 
$R=7\arcmin$ (Ly\,4), $R=9\arcmin$ (Tr\,23), $R=6.1\arcmin$ (Ly\,9), $R=14\arcmin$ (Tr\,26), and
$R=5.2\arcmin$ (Cz\,37).
\end{list}
\end{table*}

Stars that remain in the CMD after application of the FS decontamination are in cells where the stellar
density presents a clear excess over the field. Consequently, they have a significant probability of being 
cluster members. In crowded field regions, however, FS density at faint magnitudes (which presents an
exponential growth with magnitude - Fig.~\ref{fig9}) may be equal or even larger than that measured for 
the cluster. In such cases the present FS decontamination algorithm naturally truncates the MS at about the 
same level as that set by the 2MASS completeness limits (Table~\ref{tab2}).

\subsection{Field-star decontaminated CMDs}
\label{Decont_CMDs}

FS-decontaminated CMDs of the present objects are shown in the bottom panels of Figs.~\ref{fig5}-\ref{fig7}, 
where we also include the respective CM filters (Sect.~\ref{CMF}). As expected, most of the red and faint
stars were indicated as background by the decontamination algorithm. In all cases, the remaining stars in
the CMDs populate up sequences of typical intermediate-age OCs, with marked TO and GB. NGC\,5715 is a
clear cluster with a populated MS spanning $\approx3$\,mag and a conspicuous clump. Tr\,23 is a
cluster (as concluded also by \citealt{CJCM06}) as denoted by the populated MS and a conspicuous clump,
although residual FS contamination is more important than in NGC\,5715. Ly\,9 exhibits prominent MS and
clump indicative of a cluster (Sect.~\ref{GC_Ly9}). Tr\,26 is a cluster with a rather populated MS
considerably bluer than the bulge star contamination. Ly\,4 is a more complex object whose FS-decontaminated
CMD suggests a MS but residual contamination is important throughout the CMD. However, RDPs strongly
indicates a cluster nature (Sect.~\ref{RDPs} and Fig.~\ref{fig10}). Cz\,37 presents sequences considerably
bluer than the field which favours the analysis. It displays about 2\,mag of the MS and an apparently
contaminated clump. However, RDPs of MS and giant clump taken separately (Sect.~\ref{GC_Cz37} and
Fig.~\ref{fig10}) show that we are dealing with a cluster.

Residual bulge star contamination occurs in the CMDs of the 2 most centrally located objects, Tr\,26 
(bottom panel of Fig.~\ref{fig6}) and Cz\,37 (Fig.~\ref{fig7}). However, the fractions of non-subtracted 
bulge stars with respect to the observed ones are $\approx7\%$ and $\approx26\%$ for $R\leq3\arcmin$,
respectively, consistent with the subtraction efficiencies (Table~\ref{tab3}).

As described above (Sect.~\ref{Descr_FSD}), the FS decontamination algorithm considers only magnitude 
and colour properties of stars to compute the expected number of FSs in CMD cells. As a consequence, 
stars in a given CMD cell (that can have any radial coordinate) have the same probability of being deleted. 
However, star clusters have stellar densities higher at the core than the halo. Thus, in the case of
real star clusters, random subtraction of stars in cells tend to artificially change the intrinsic 
statistics of the radial distribution of stars. In other words, FS-decontaminated photometry retains 
only the colour/magnitude information, but not the radial one. In this sense, we used the FS 
decontamination algorithm only to uncover intrinsic CMD morphologies.

\subsection{Potential limitations}
\label{Pot_Lim}

In principle, the present FS decontamination method is capable of taking into account local density
variations of field and cluster star sequences in CMDs (Sect.~\ref{Descr_FSD}). To reach that goal,
however, it relies on some uniformity - at least to the CMD cell scale - of statistical properties
(e.g. colour and magnitude distributions), both in the offset field and cluster regions. Such premises
may introduce some limitations to the method, basically related to photometry. Some of these are discussed
below.

\begin{itemize}

\item[\em (a)] {\em Significant differential absorption.} Consider a CMD cell whose colour sides have 
dimensions $\Delta\jh=\Delta\jk=0.2$. If the difference in absorption between cluster and offset field 
stars for a given cell is larger than $A_V\approx2$\,mag, the difference in colour (e.g. \jh) would be 
as large as the cell dimension. In that case the algorithm would assume wrong types (and possibly number)
of stars as FS to be subtracted from the cell.

\item[\em (b)] {\em Large photometric uncertainties.} Because the algorithm uses explicitly photometric
errors (in terms of Gaussian probability distribution) to compute cell number density, exceedingly
large uncertainties would result both in high fractions of cells with fractional numbers of stars
(followed by low subtraction efficiency) and smeared CMDs.

\item[\em (c)] {\em Small number of cluster stars.} CMD cells of (intrinsically) poorly-populated
clusters inevitably suffer from small-number statistics that would result in low subtraction efficiencies.
In addition, the effective number of stars above the background in photometry-limited surveys falls with
distance from the sun as $N^\star_{eff}\propto1/\ds$ (\citealt{DetAnalOCs}), which decreases star-count
statistics of distant clusters.

\item[\em (d)] {\em Distant and central clusters.} The main problem in these cases is the exponential growth
of FS contamination with (apparent) magnitude (Sect.~\ref{Applic}). The fraction of cluster stars in
very distant and central objects may drop to a few percent, much below the $1\sigma$ fluctuation level,
especially at the 2MASS completeness limits (Sect.~\ref{2mass}).

\item[\em (e)] {\em Crowding in central parts.} Rich clusters, and to some extent distant ones, may suffer
from stellar crowding especially in their central regions. Angular separations smaller than
$\approx1\arcsec$ cannot be resolved by 2MASS. Since the offset field is less affected by crowding than
clusters, FS contamination, especially for faint stars, would be overestimated.

\item[\em (f)] {\em CMD cell dimension.} Large CMD cells might be used to minimize effects {\em (a)} --
{\em (c)}, but the consequence would be a degraded CMD resolution (Sect.~\ref{Descr_FSD}).

One possible way to circumvent part of the above limitations would be to apply the FS decontamination 
algorithm on CCD photometry. Their small photometric uncertainties probably would produce less fractional
numbers in cells and, thus, higher subtraction efficiencies. However, for any quantitative FS decontamination
algorithm to be efficient and statistically representative, wide surrounding fields such as those provided
by 2MASS, are required. Wide fields in the optical are usually not available. The present objects do
not suffer critically from the effects above. 

\end{itemize}

\subsection{Cluster age, reddening and distance from the Sun derived from FS-decontaminated data}
\label{age}

Cluster age is derived with solar-metallicity Padova isochrones (\citealt{Girardi2002}) computed 
with the 2MASS \jj, \hh\ and \ks\ filters\footnote{\em
http://pleiadi.pd.astro.it/isoc$\_$photsys.01/isoc$\_$photsys.01.html }. 2MASS transmission filters
produced isochrones very similar to the Johnson-Kron-Cousins (e.g. \citealt{BesBret88}) ones,
with differences of at most 0.01 in \jh\ (\citealt{TheoretIsoc}).

The FS decontaminated CMD morphologies (bottom panels of Figs.~\ref{fig5} to \ref{fig7}) provide enough 
constraints to derive reliable cluster ages. We derive ages in the range 0.6 -- 1.3\,Gyr, with Cz\,37 
being the youngest and Ly\,4 the oldest cluster (col.~9 of Table~\ref{tab1}). Reddening values are in the 
range $\ejh=0.11 - 0.37$ which convert to $\ebv=0.35 - 1.18$ (col.~10). Distances from the Sun are in the 
range $\ds=1.0 - 1.7$\,kpc (col.~11). Age-solutions are plotted in the bottom panels of Figs.~\ref{fig5} 
to \ref{fig7}, superimposed on the FS-decontaminated CMDs. 

The present age of Tr\,23 agrees with that of \citet{CJCM06}. However, they derive larger
reddening ($\Delta\ebv\approx+0.3$) and distance from the Sun ($\Delta\ds\approx+0.5$\,kpc).
With respect to Tr\,26, \citet{Kharchenko05} found age and reddening $\sim1/3$ of the present
ones, and put the cluster $\sim3$ times more distant. 

With the recently derived value of the Sun's distance to the Galactic centre $R_O=7.2$\,kpc (based 
on updated parameters of globular clusters - \citealt{GCProp}), we conclude that the present OCs 
are located $\approx0.9 - 1.6$\,kpc inside the Solar circle (col.~12 of Table~\ref{tab1}). Cz\,37 
is the innermost one, with Galactocentric distance $\dgc\approx5.6$\,kpc.

\section{Structural parameters}
\label{struc}

Structural parameters are derived by means of RDPs, defined as the projected radial distribution of
the number-density of stars around the cluster centre. RDPs are built with stars selected after
applying the respective CM filter (bottom panels of Figs.~\ref{fig5} - \ref{fig7}) to the observed 
photometry. 

\subsection{Colour-magnitude filters}
\label{CMF}

CM filters were previously applied in the structural analysis of the OCs M\,67 (\citealt{M67}), NGC\,3680
(\citealt{N3680}), NGC\,188 (\citealt{N188}), NGC\,6611 (\citealt{N6611}), NGC\,4755 (\citealt{N4755}), 
M\,52 and NGC\,3960 (\citealt{M52N3960}) and the faint OCs BH\,63, Lyng\aa\,2, Lyng\aa\,12 and King\,20
(\citealt{FaintOCs}). They are used only to discard  stars with colours compatible with those of 
the foreground/background field. They should be wide enough to accommodate cluster MS and evolved 
stars colour distributions, allowing as well for the respective $1\sigma$ uncertainties. Contrarily to CMD 
FS decontamination (Sect.~\ref{FSD}), residual field stars with colours similar to those of the cluster
are expected to permeate the CM filter. This residual contamination is statistically taken into account 
by comparisons with the offset field. Hence the need for wide surrounding fields such as those provided by 
2MASS. CM filter widths account for dynamical evolution-related effects, such as enhanced fractions of 
binaries (and other multiple systems) towards the central parts of clusters, since such systems tend to 
produce a widening in the MS. Effects on CM-filter shape due to binary-induced MS widening and 2MASS 
photometric uncertainties have been studied in the old OC NGC\,188 (\citealt{N188}).

It is important to make clear that, in the context of structural and MF analyses, FS-decontaminated CMDs 
(bottom panels of Figs.~\ref{fig5}-\ref{fig7}) are used exclusively to define the shape of CM-filters. 
They are designed to contain intrinsic cluster sequences (including as well MS-widening 
evolutionary effects). RDPs and MFs employ the CM-filtered observed photometry.

\subsection{Radial Density Profiles}
\label{RDPs}

To avoid oversampling near the centre and undersampling at large radii, RDPs are built by counting stars 
in rings of increasing width with distance to the centre. The number and width of rings can be adjusted 
so that the resulting RDPs present good spatial resolution with moderate $1\sigma$ Poisson errors. The 
residual background level of each RDP corresponds to the average number of (CM-filtered) stars measured 
in the comparison field (Table~\ref{tab4}). 

Cluster limiting radius and uncertainty are estimated by visually comparing the RDP level (taking into
account fluctuations) with the background. In this sense, \rl\ corresponds to the distance from the 
cluster centre where RDP and background become statistically indistinguishable from each other (e.g.
\citealt{DetAnalOCs}, and references therein). For practical purposes, most of the cluster stars can be 
considered to be contained within $\rl$. We remark that the limiting radius should not be mistaken for the 
tidal radius. For instance, in populous and relatively high Galactic latitude OCs such as M\,26, M\,67, 
NGC\,188 and NGC\,2477, limiting radii are a factor $\sim0.5 - 0.7$ of the respective tidal radii 
(\citealt{DetAnalOCs}). The limiting radii of the present objects are given in col.~7 of Table~\ref{tab4}. 
It is worth remarking that tidal radii are derived from fits of King profile to RDPs, which depend on wide 
surrounding fields and adequate Poisson errors.

\begin{figure}
\resizebox{\hsize}{!}{\includegraphics{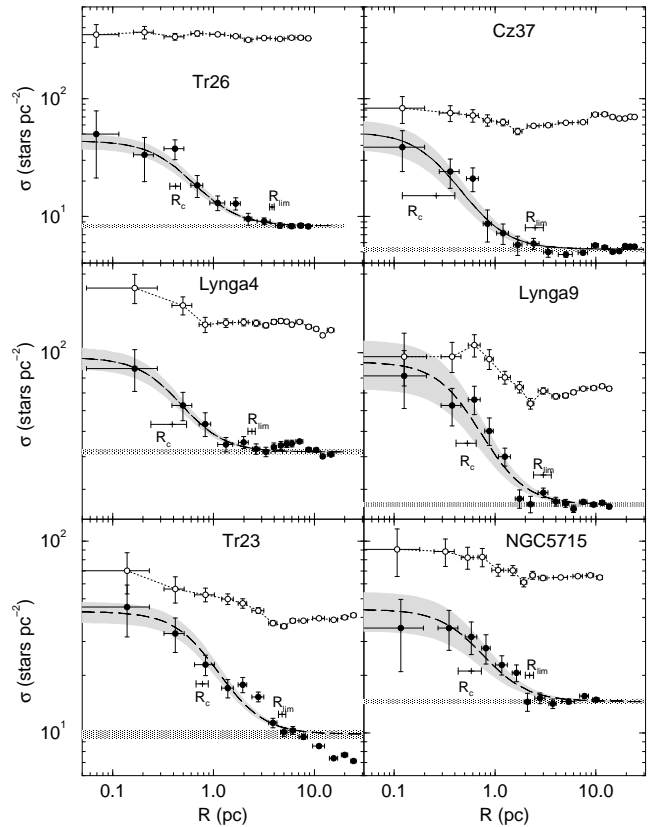}}
\caption[]{Radial stellar density profiles. Filled circles: colour-magnitude filtered RDPs. Dashed line:
best-fitting two-parameter King profile. Horizontal shaded region: residual stellar background level 
measured in the comparison field. Core and limiting radii are indicated. Gray regions: standard deviation 
of the King fit. The striking effects of the background contamination, that produce shallow and/or disturbed
profiles, are reflected in the low-contrast observed RDPs (empty circles).}
\label{fig10}
\end{figure}

\begin{figure}
\resizebox{\hsize}{!}{\includegraphics{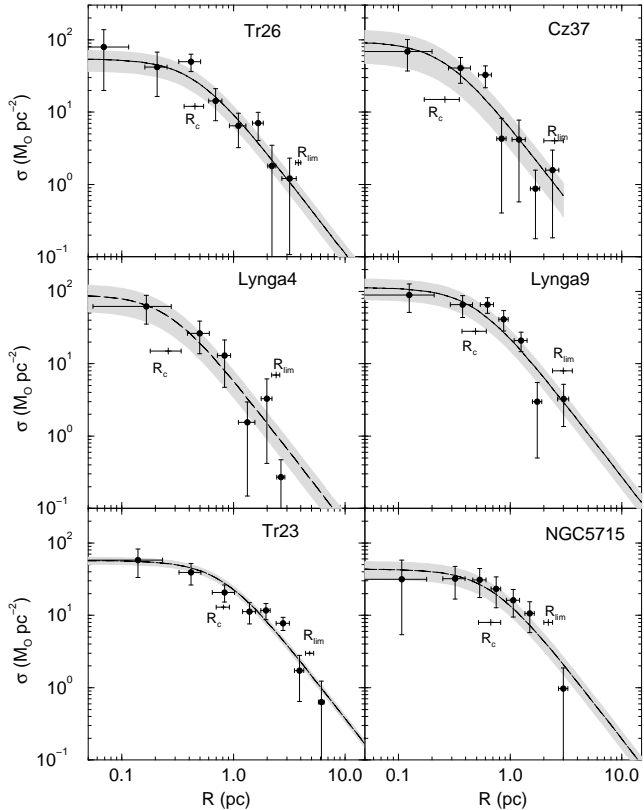}}
\caption[]{Similarly to RDPs (Fig.~\ref{fig10}), background-subtracted MDPs (filled circles) are well 
represented by two-parameter King profiles (dashed lines) with relatively small uncertainties (shaded 
regions). MDPs core radii agree at $1\sigma$ with those derived from RDPs (Table~\ref{tab4}).}
\label{fig11}
\end{figure}

\begin{table*}
\caption[]{Structural parameters from CM-filtered photometry}
\label{tab4}
\renewcommand{\tabcolsep}{2.7mm}
\renewcommand{\arraystretch}{1.3}
\begin{tabular}{lcccccccccc}
\hline\hline
&&&\multicolumn{5}{c}{RDP}&&\multicolumn{2}{c}{MDP}\\
\cline{4-8}\cline{10-11}
Cluster&$1\arcmin$&&$\sigma_{bg}$&$\sigma_{0K}$&$\delta_c$&\rc&\rl&&$\sigma_{0K}$&\rc \\
       &(pc)&&$\rm(stars\,pc^{-2})$&$\rm(stars\,pc^{-2})$&&(pc)&(pc)&
       &$\rm(\ms\,pc^{-2})$&(pc)\\
(1)&(2)&&(3)&(4)&(5)&(6)&(7)&&(8)&(9)\\
\hline
NGC\,5715&0.427&&$14.6\pm0.2$&$30\pm10$&$2.1\pm0.7$&$0.58\pm0.15$&$2.2\pm0.2$&&$44\pm12$&$0.67\pm0.15$\\
Ly\,4    &0.332&&$41.6\pm0.2$&$54\pm15$&$1.3\pm0.4$&$0.39\pm0.15$ &$2.4\pm0.2$&&$90\pm36$&$0.26\pm0.08$\\
Tr\,23   &0.553&&$9.9\pm0.4$&$33\pm5$&$3.3\pm0.5$&$0.78\pm0.11$ &$4.8\pm0.4$&&$57\pm6$&$0.81\pm0.11$\\
Ly\,9    &0.499&&$26.2\pm0.2$&$66\pm19$&$2.5\pm0.7$ &$0.53\pm0.12$&$3.0\pm0.6$&&$113\pm36$&$0.49\pm0.12$\\
Tr\,26   &0.276&&$8.3\pm0.2$&$36\pm6$&$4.4\pm0.7$ &$0.42\pm0.05$&$4.0\pm0.2$&&$54\pm17$&$0.45\pm0.09$\\
Cz\,37   &0.479&&$5.3\pm0.1$&$46\pm15$&$8.7\pm2.8$ &$0.26\pm0.14$&$2.5\pm0.5$&&$94\pm25$&$0.26\pm0.09$\\
\hline
\end{tabular}
\begin{list}{Table Notes.}
\item Col.~2: arcmin to parsec scale. King profile is expressed as 
$\sigma(R)=\sigma_{bg}+\sigma_{0K}/(1+(R/R_{\rm core})^2)$. To minimize degrees of freedom
in RDP fits, $\sigma_{bg}$ was kept fixed (measured in the respective comparison fields) while
$\sigma_{0K}$ and \rc\ were allowed to vary. MDPs are background subtracted profiles. Col.~5:
cluster/background density contrast ($\delta_c=\sigma_{0K}/\sigma_{bg}$), measured in CM-filtered
RDPs.
\end{list}
\end{table*}

Structural parameters are derived by fitting CM-filtered RDPs with the two-parameter \citet{King1966a}
profile, which describes the central and intermediate regions of normal clusters (\citealt{King1966b};
\citealt{TKD95}). Fits were performed with a non-linear least-squares fit routine that uses errors 
as weights. To minimize degrees of freedom in fits, the background level ($\sigma_{bg}$ - col.~3 of
Table~\ref{tab4}) was kept constant, corresponding to the residual values measured in the comparison 
fields (Sect.~\ref{FSD}). King fit parameters are the residual central density of stars ($\sigma_{0K}$) 
and core radius (\rc). Fit parameters are given in cols.~4 and 6 of Table~\ref{tab4}, and the
best-fitting solutions are superimposed on the CM-filtered RDPs (Fig.~\ref{fig10}). We quantify the
background contamination by the density contrast parameter $\delta_c=\sigma_{0K}/\sigma_{bg}$ (col.~5).
Since $\delta_c$ is measured in CM-filtered RDPs, it does not necessarily correspond to the visual
contrast produced by observed stellar distributions in XDSS images (Figs.~\ref{fig1} - \ref{fig3}). Tr\,26,
for instance, presents a very-low contrast both in the XDSS B image (Fig.~\ref{fig1}) and observed
RDP (Fig.~\ref{fig10}), where $\delta_c\approx0.1$. On the other hand, because most of the non-cluster
stars have been excluded by the CM filter, the corresponding RDP presents a higher density contrast,
$\delta_c\approx4.4$.

Probably because of different methods and data sets, the present values of \rc\ and \rl\ for Tr\,26 
correspond to factors of $0.4$ and $1.5$ of those given in (\citealt{Kharchenko05}). The difference,
especially in \rc, may be attributed to their brighter limits (\citealt{KPRSS04}) producing shallower 
profiles for this bulge-contaminated cluster. The average values of core and limiting radii for the
present objects,
$\langle\rc\rangle=0.47\pm0.14$\,pc and $\langle\rl\rangle=3.0\pm0.9$\,pc, are smaller than the
corresponding ones measured in the OC sample of \citet{Schi2006}, $\langle\rc\rangle=2.0\pm1.3$\,pc 
and $\langle\rl\rangle=5.0\pm3.2$\,pc.

Fig.~\ref{fig10} shows RDPs of MS/evolved stars of the present objects. For absolute comparison between
clusters the radius scale was converted to parsecs and the number-density of stars to $\rm stars\,pc^{-2}$\
using distances derived in Sect.~\ref{age}. RDPs built with observed photometry are also shown in
Fig.~\ref{fig10}. Clearly, CM-filtered profiles present less fluctuations and probe deeper into cluster 
structure than observed RDPs. In particular, observed profiles tend to underestimate cluster extension. 
Besides, CM filters were essential to unveil the cluster centroid (Sect.~\ref{Target_OCs}). Note 
that the RDP of Tr\,23 keeps decreasing for $R\ga\rl$, an effect produced by the variable north-south 
absorption (Sect.~\ref{DiffRedInTr23}). 

Within uncertainties, King profile provides a good analytical representation of the stellar RDPs of
the six objects, from the outer parts to the core. Since it follows from an isothermal (virialized) 
sphere, the close similarity of a cluster's stellar RDP with a King profile suggests that the internal 
structure (particularly the core) has reached some significant level of energy equipartition, which is
consistent with the ages derived for these objects (Sect.~\ref{CWODS}).

Striking differences between observed and CM-filtered RDPs show up in Fig.~\ref{fig10}. It is remarkable that 
in all cases, CM-filtered RDPs present high contrast with respect to the background and are well fitted 
by King profiles. On the other hand, observed RDPs are in general shallow and irregular, clearly not 
typical of star clusters. The photometry from which the observed RDPs were built is dominated by FS 
contamination, in general by a fraction larger than 90\% (Table~\ref{tab3}). Accordingly, the fraction 
of cluster stars, with respect to that of the background, is at the $1\sigma$ level. In this sense, it 
should be expected that without FS decontamination (for CMD morphology) and CM-filters (for intrinsic 
RDPs), simple qualitative comparisons of candidate CMDs with offset fields can be misleading, especially 
for clusters projected in central parts of the Galaxy.

\subsection{Mass Density Profiles}
\label{MDPs}

To complete the structural description of the objects we take the mass-luminosity (ML) relation derived
from isochrone fits (Sect.~\ref{age}) to build statistical mass-density profiles. We 
follow the same systematics as that used to build RDPs. Instead of  computing the
number-density of stars in rings, we now assign each star a mass according to the respective 
ML relation. MDPs are produced by subtracting from the mass density in each ring that
measured in the comparison field. They are shown in Fig.~\ref{fig11}, together with the respective
King fits. Likewise RDPs, MDPs are well described by King profiles. Core radii derived
from MDPs (col.~9 of Table~\ref{tab4}) agree, at $1\,\sigma$, with RDP ones (col.~6).

\subsection{The case of Lyng\aa\,9}
\label{GC_Ly9}

Lyng\aa\,9 was argued to be an asterism by \citet{CJE05}. Based on a comparison with an offset field and 
applying a visual method to delete field star contamination, they eliminated most MS stars, except for 
some at the top of the MS, and a group of bright, red stars that resembled a giant clump. They concluded 
that Ly\,9 was not a star cluster, and that the clump could consist of a random distribution of early-type 
stars behind of or embedded in a hypothetical obscuring cloud.

However, the present FS decontamination algorithm (Sect.~\ref{FSD}) showed statistically
significant star excesses over the field that populate the giant clump and about 3 magnitudes
of the MS (bottom panel of Fig.~\ref{fig6}). In addition, CM-filtered photometry (Sect.~\ref{struc})
produced high contrast King-like RDP (Fig.~\ref{fig10}) and MDPs (Fig.~\ref{fig11}).

\begin{figure}
\resizebox{\hsize}{!}{\includegraphics{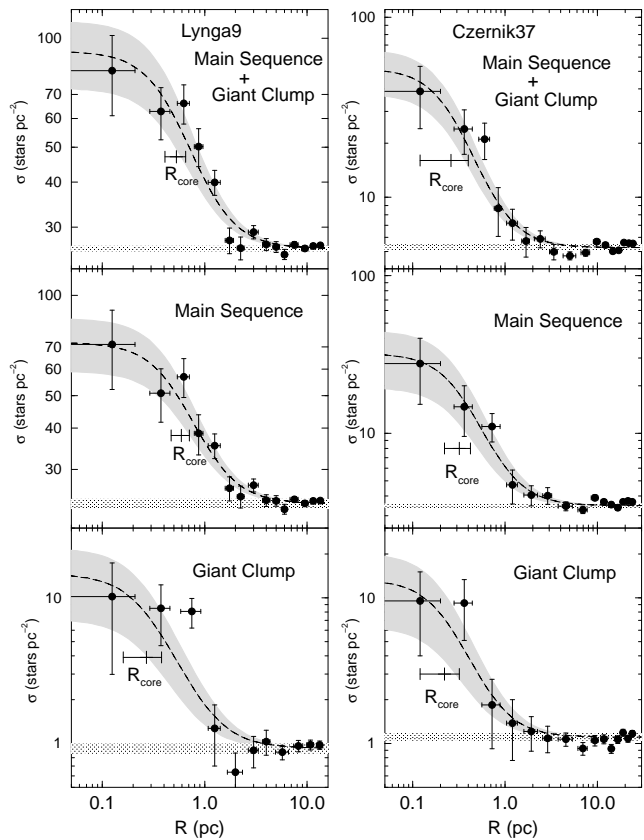}}
\caption[]{RDPs of  Ly\,9 (left panels) and Cz\,37 (right panels) built
with CM-filtered photometry for different magnitude ranges. Top panels: full magnitude range,
i.e. MS $+$ giant clump stars. Middle panels: MS stars. Bottom panels: Giant clump stars. Symbols
as in Fig.~\ref{fig10}.}
\label{fig12}
\end{figure}

To further explore the scenario emerging from the FS-decontaminated CMD, we build RDPs for MS and
clump stars separately. If Ly\,9 is a star cluster, RDPs of both star subsamples should present
similar, King-like, features. As before (Sect.~\ref{RDPs}), RDPs are built with CM-filtered
photometry. The resulting MS and clump RDPs are shown in Fig.~\ref{fig12} (middle and bottom panels,
respectively), where we show as reference the combined MS $+$ giant clump RDP (top panel) taken from
Fig.~\ref{fig10}. Within uncertainties, both RDPs can be reproduced by similar King profiles,
respectively with $\rc=0.59\pm0.12$\,pc and $\rc=0.27\pm0.12$\,pc. The smaller core radius of the
clump RDP, as compared to that of the MS, is consistent with dynamical evolution affecting Ly\,9
(Sect.~\ref{CWODS}), where giants are expected to be more concentrated in the cluster's central parts.

Considering that Ly\,9 presents a populated MS and that its RDP, as well as that of the clump,
are similar and follow a King profile, we conclude that we are dealing with an open cluster.

\subsection{The case of Czernik\,37}
\label{GC_Cz37}

\citet{CJE05}, based on $(B-V)$ and
$(V-I)$ CMDs of the object (however with no offset field for comparison) suggested that it might be
a star cluster superimposed on the Galactic bulge population. It is the least-populated object of the
present sample (Table~\ref{tab5}), with a FS-decontaminated CMD featuring about 2 magnitudes of the MS,
a relatively conspicuous clump, and significant bulge star contamination (Fig.~\ref{fig7}). However,
its CM-filtered RDP is highly contrasted with the background and follows closely a King profile
(Fig.~\ref{fig10}).

We further check the OC nature of Cz\,37 with RDPs built separately with (CM-filtered)
MS and clump stars. The resulting MS and clump profiles (right panels of Fig.~\ref{fig12}) are similar
to each other, slightly more concentrated for the clump than the MS. They follow King laws with core radii
$\rc=0.32\pm0.10$\,pc and $\rc=0.22\pm0.10$\,pc, respectively for the MS and clump RDPs. 
 Accordingly, Cz\,37 is confirmed to be an open cluster.

\subsection{Model OCs projected on varying backgrounds}
\label{Simul}

Different luminosity functions detected in central parts of most star clusters, as compared to the 
field, may provide a means from which these objects can identified in plate materials. However, 
in the context of the present work, structural parameters are derived from number-density profiles
that consider only the number of stars in rings, regardless of magnitude. In short, the issue here 
is not star cluster identification, but to measure core and limiting radii from RDPs of objects
subject to different amounts of background contamination.

To further explore the OC structural description we test the effect of a varying background level
on measurements of cluster core and limiting radii. Since determination of \rl\ depends directly on 
the ability to see where the RDPs become indistinguishable from the background level, it is expected
that measurements of non-populous OCs, especially those projected against central parts of the
Galaxy, will be underestimated. 

\begin{figure}
\caption[]{Visual contrast produced by a model (King profile) OC described by $\rc=0.4$\,pc, $\rl=4.0$\,pc
and 350 member stars projected against backgrounds with varying density levels
($\delta_c=\sigma_{0K}/\sigma_{bg}$). Top-left panel: no background. Top-right: $\delta_c=10$.
Bottom-left: $\delta_c=3.3$. Bottom-right: $\delta_c=2$.}
\label{fig13}
\end{figure}

\begin{figure}
\resizebox{\hsize}{!}{\includegraphics{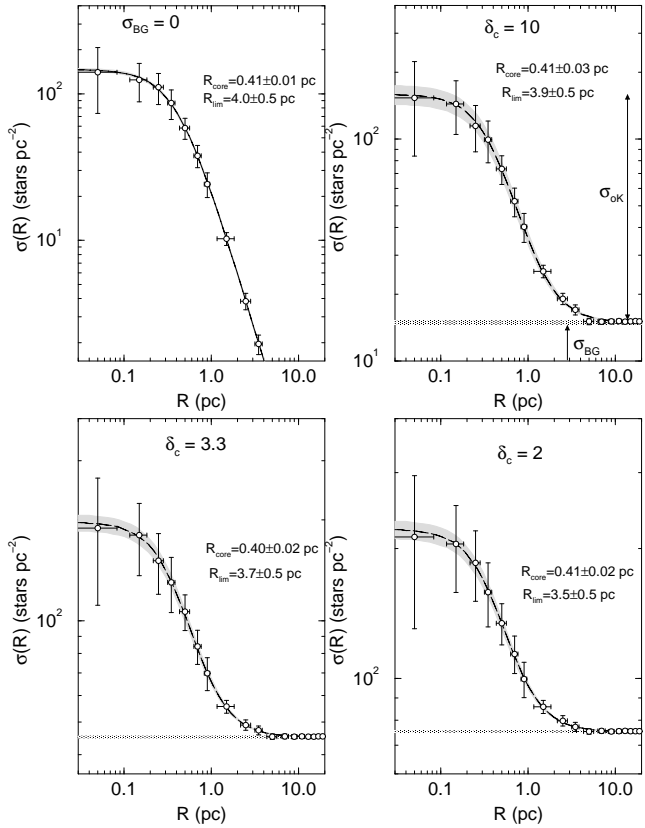}}
\caption[]{RDPs of the model OCs shown in Fig.~\ref{fig13}. Limiting radius tends to be underestimated
as the density contrast parameter drops. Panels follow the same order of $\delta_c$ as in Fig.~\ref{fig13}.
Core (from King fit) and limiting radii (from RDPs) are indicated. For visualization purposes, vertical
scales are different in each panel.}
\label{fig14}
\end{figure}

King-like OCs are simulated with (projected) stellar radial number-density profiles given by
$\sigma(R)=\sigma_{bg}+\sigma_{0K}/(1+(R/R_{\rm core})^2)$. We define the cluster/background contrast
parameter as $\delta_c=\sigma_{0K}/\sigma_{bg}$. Model OCs contain 350 equal-mass stars distributed
in a region with a structure described by $\rc=0.4$\,pc and $\rl=4.0$\,pc, representative values of 
the OCs dealt with in this work (Table~\ref{tab4}).  (x,y) coordinates of the test-stars are
randomly selected with a probability proportional to King profile's number density at position 
$R=\sqrt{x^2+y^2}$, in steps of $\Delta R=0.01$\,pc, in order to preserve the spatial resolution of 
the analytical profile. To avoid small-number statistics at inner regions (the small ring-areas may 
contain fractions of stars), we run 100 simulations and take the average probability at each position 
to build the final RDPs.

The spatial distribution of stars of the background-free OC is shown in Fig.~\ref{fig13}
(top-left panel), where the core and halo subsystems are clearly visible. To this OC we add backgrounds 
with $\delta_c=$ 10 (top-right), 3.3 (bottom-left) and 2 (bottom-right). As expected, only the
(equal-mass star) core detaches from the stellar distributions with increasing background level.
The picture does not change much in real star clusters, since bright stars in these objects tend
to be concentrated in or near the core. As a consequence of low-surface brightness, the stellar
distribution of the cluster halo, especially in the outer parts, is not much different from 
that of the field.

The corresponding model-OC RDPs are shown in Fig.~\ref{fig14}, together with the King fits and
background levels. RDPs are built with a similar ring-width distribution as that used for the RDPs
and MDPs in Figs.~\ref{fig10} and \ref{fig13}, respectively. As expected, core radii derived from
King fits do not depend on $\delta_c$. Limiting radii, on the other hand decrease with increasing
$\delta_c$, changing from $\rl=4.0\pm0.5$ (background-free) to $\rl=3.5\pm0.5$ for $\delta_c=2$, a 
$\sim10\%$ factor.

As a caveat we note that the backgrounds used in the simulations are spatially uniform, subject
only to $1\sigma$ Poisson fluctuations in RDPs. Galactic backgrounds present higher spatial
fluctuations (e.g. Figs.~\ref{fig1} to \ref{fig3}) associated with e.g. differential absorption
and local FS enhancements that tend to produce additional decrease in limiting radii. 
Thus, we expect to measure limiting radii about 10-20\% smaller than intrinsic ones
in OCs containing a few hundred stars projected against central parts of the Galaxy.
Obviously, OCs less-populous than those simulated here are expected to be significantly more 
affected by low contrast, to the point that most may not be observed at all (\citealt{DiskProp}).

\section{Mass determinations}
\label{MF}

\subsection{Mass function slopes}

In previous sections we presented strong evidence, in terms of CMDs, RDPs and MDPs,
that the 6 objects dealt with in this work are intermediate-age OCs. Based on this we 
build MFs and compute stellar masses using CM-filtered photometry (Sect.~\ref{struc}),
that increases statistical significance and cluster membership probability.

The methods presented in \citet{DetAnalOCs} (and
references therein) are used to derive MFs, $\left(\phi(m)=\frac{dN}{dm}\right)$. We build them 
using the three 2MASS bands separately, and the mass-luminosity relations obtained from the respective Padova 
isochrones and distances from the Sun (Sect.~\ref{2mass}).  The
effective magnitude range over which MFs are computed is that where clusters present an
excess of stars with respect to the comparison field. In all cases it begins right below the TO and
ends at a faint-magnitude limit brighter than that stipulated by the 2MASS completeness limit
(Sect.~\ref{2mass}). The effective magnitude ranges of the present OCs are shown as shaded areas
in the bottom panels of Figs.~\ref{fig5} to \ref{fig7}; the corresponding stellar mass ranges are
given in col.~4 of Table~\ref{tab5}. Further details on MF construction are given in \citet{FaintOCs}.

\begin{table*}
\caption[]{Parameters related to mass functions and dynamical states}
\label{tab5}
\renewcommand{\tabcolsep}{1.3mm}
\renewcommand{\arraystretch}{1.2}
\begin{tabular}{cccccccccccccc}
\hline\hline
&\multicolumn{2}{c}{Evolved}&&\multicolumn{4}{c}{Observed MS}&&
\multicolumn{5}{c}{$\rm Evolved\ +\ Extrapolated\ MS\ $}\\
\cline{2-3}\cline{5-8}\cline{10-14}
Cluster&$N^*$&$m$&&$\Delta m$&$\chi$&$N^*$&$M_{obs}$&&$N^*$&$M_{OC}$&$\sigma$&$\rho$&$\tau$\\
&(stars)&(\ms)&&(\ms)&&(stars)&(\ms)&& ($10^2$stars)&($10^2\ms$)&($\rm \ms\,pc^{-2}$)&($\rm \ms\,pc^{-3}$)\\
 (1)  & (2)   & (3)  &&(4)     & (5)    & (6)   & (7)    && (8)   &(9) &(10) & (11)& (12)\\
\hline
NGC\,5715& $8\pm2$&$18\pm6$ &&1.1--2.2&$1.3\pm0.5$& $93\pm23$&$145\pm20$
&&$20\pm15$&$7.3\pm3.1$&$48\pm21$&$16\pm7$&$35\pm25$\\
Ly\,4& $3\pm1$&$7\pm2$ &&1.0--1.9&$1.3\pm0.2$& $105\pm40$&$130\pm30$
&&$22\pm16$&$7.6\pm3.0$&$42\pm17$&$13\pm5$&$41\pm28$\\ 
Tr\,23& $16\pm9$&$35\pm20$ &&1.0--2.1&$1.4\pm0.6$& $300\pm40$&$483\pm42$
&&$86\pm67$&$31\pm13$&$43\pm18$&$6.7\pm2.8$&$4.9\pm3.5$\\
Ly\,9& $37\pm7$&$73\pm15$ &&1.1--1.9&$1.1\pm0.7$& $124\pm32$&$213\pm37$
&&$26\pm22$&$10\pm5$&$37\pm17$&$9.2\pm4.1$&$18\pm14$\\ 
Tr\,26& $21\pm5$&$42\pm10$ &&1.1--2.3&$1.7\pm0.6$& $100\pm31$&$150\pm24$
&&$24\pm18$&$8.7\pm3.5$&$17\pm7$&$3.3\pm1.3$&$15\pm10$\\
Cz\,37& $17\pm5$&$44\pm13$ &&1.8--2.5&$-1.1\pm0.9$& $41\pm14$&$89\pm19$
&&$1.4\pm0.7$&$2.1\pm1.0$&$11\pm5$&$3.2\pm1.6$&$196\pm88$\\ 
\hline\hline
\end{tabular}
\begin{list}{Table Notes.}
\item Col.~4: effective mass range of the observed MS. Col.~7: stellar mass of the observed MS.
Col.~9: mass extrapolated to 0.08\,\ms. Col.~12: dynamical-evolution parameter
$\tau={\rm age}/t_{\rm rel}$.
\end{list}
\end{table*}

\begin{figure} 
\resizebox{\hsize}{!}{\includegraphics{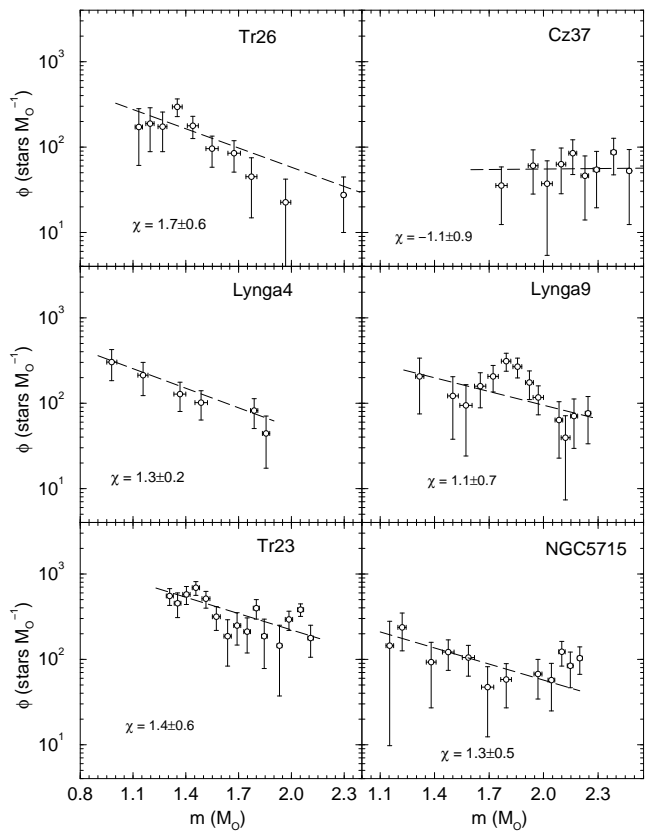}}
\caption[]{2MASS mass functions fitted with $\phi(m)\propto m^{-(1+\chi)}$. Except for Cz\,37, the
remaining OCs have MF slopes similar to Salpeter's IMF.}
\label{fig15}
\end{figure}

The resulting MFs cover significant mass ranges, typically from 1 to 2.5\ms. They are shown in 
Fig.~\ref{fig15}, where fits with the function $\phi(m)\propto m^{-(1+\chi)}$ are included; MF
slopes are given in col.~5 of Table~\ref{tab5}. Error bars basically reflect the significant number
of stars present in the MS. Within uncertainties, the MFs of NGC\,5715, Ly\,4, Tr\,23, Ly\,9 and
Tr\,26 have slopes similar to that of \citet{Salpeter55} IMF ($\chi=1.35$). Cz\,37, on the other
hand, presents a flat MF which may result from an advanced dynamical state (Sect.~\ref{CWODS}).

\subsection{Cluster mass}
\label{Cl_mass}

Table~\ref{tab5} gives parameters of the target clusters measured in the CMDs and derived from 
MFs. The number of evolved stars (col.~2) was obtained by integration of the (background-subtracted)
CM-filtered luminosity function for stars brighter than the TO. Multiplying this
number by the stellar mass at the TO yields an estimate of the mass stored in evolved stars (col.~3).
The number and mass of the observed MS stars (cols.~6 and 7, respectively) were derived by
integrating the MFs over the effective MS mass ranges (col.~4).

To estimate the total stellar mass we extrapolate the observed MFs down to the H-burning mass limit
($0.08\,\ms$). For
masses below the present detection threshold (Table~\ref{tab4}) we rely the extrapolation on
\citet{Kroupa2001} universal Initial Mass Function (IMF), in which $\chi=0.3\pm0.5$ for the range
$0.08\leq m(\ms)\leq0.5$ and $\chi=1.3\pm0.3$ for $0.5\leq m(\ms)\leq1.0$. In the cases where the
present MF slopes are flatter than or similar (within uncertainties) to Kroupa's, we adopt the measured
values of $\chi$. The total (extrapolated MS $+$ evolved) values of number, mass, projected and volume
densities are given in cols.~8 to 11 of Table~\ref{tab5}.

The number of MS and evolved member stars ranges from $\approx58$ to $\approx316$. As already indicated
by the small number of probable member stars present in the FS decontaminated CMD (Fig.~\ref{fig7}), 
the least populous OC is Cz\,37. The corresponding mass of the MS and evolved stars is in the range
$130\la M_{obs}(\ms)\la520$, while the extrapolated masses are a factor $\sim4.5 - 6$ times larger than the
observed ones, except for Cz\,37, which is only about twice as large, again consistent with its
CMD morphology.

As a caveat we note that the total mass estimates should be taken as upper limits, since because
of dynamical evolution, significant fractions of the low-mass content may have been lost to the
field, especially in the case of Cz\,37 (Sect.~\ref{CWODS}), as reflected in the flat MF.

\section{Discussion}
\label{CWODS}

Star clusters form in collapsing molecular clouds in which variable fractions (10 -- 60\%) of the 
parent gas are converted into stars. They remain embedded in the clouds for about 2-5\,Myr (e.g. 
\citealt{LL2003}), and their dynamical state at that early phase can be described as out of virial 
equilibrium (e.g. \citealt{delaF2002}). Following the rapid expulsion of the unused gas by massive 
winds and supernovae, stars end up with excessive velocities with respect to the new, decreased, 
potential (e.g. \citealt{delaF2002}; \citealt{BK2002}). As a consequence, clusters grow in all 
scales as they reach for virialization. N-body simulations of massive star clusters (e.g. 
\citealt{GoBa06}) show that after a few $10^7$\,yr, core growth levels off as some energy equipartition 
is reached, a phase that is followed by a decrease in core size. The outer parts, on the other hand, 
keep increasing in size. At some point, the intrinsic cluster size may no longer be observable because 
of the increased volume and low number-density of cluster stars at large radii. The external parts may 
become indistinguishable from the background, particularly for clusters projected against the central 
parts of the Galaxy (e.g. \citealt{DiskProp}). Besides, interactions with the disc and the tidal pull
of the Galactic centre/bulge, as well as collisions with molecular clouds, tend to destroy poorly-populated 
OCs in a time-scale of a few $\rm10^8\,Myr$ (\citealt{Bergond2001}), especially for centrally located 
OCs. Although conceptually different, these effects combine to produce observable changes in cluster 
structural parameters. For the outer parts of poorly populated and centrally located OCs it is expected: 
{\em (i)} cluster expansion due to dynamical evolution; {\em (ii)} decrease in cluster/background contrast
especially for $R\rightarrow\rl$; {\em (iii)} enhanced disruption rates.

Empirical determinations of cluster limiting radii depend critically on RDP and background levels (and 
respective fluctuations). Thus, effects {\em (i)} and {\em (ii)} should affect more \rl\ than \rc, since 
the latter is derived by fitting an analytical function to the distribution of points provided by the RDP,
as indicated by model-cluster simulations (Sect.~\ref{Simul}). On the other hand, dynamical evolution may
accelerate because of {\em (iii)}, thus increasing the rate of large scale core-halo mass segregation
and low-mass star evaporation to the field - effects that tend to increase intrinsic cluster size and 
decrease \rc. Since mass segregation drives preferentially low-mass stars to the outer parts of clusters, 
cluster/background contrast in these regions tends to lower as clusters age. As an observational consequence, 
smaller values of limiting radii should be measured, especially for clusters in dense fields.

To disentangle these effects we employ a set of parameters that probe structure and dynamical 
state derived for a sample of nearby OCs with ages in the range $70-7\,000$\,Myr and masses within 
$400-5\,300$\,\ms, following most of the present methodology (\citealt{DetAnalOCs}). To the original
reference sample were added NGC\,6611 (\citealt{N6611}) and NGC\,4755 (\citealt{N4755}). Clusters are 
differentiated according to
total mass (smaller or larger than 1\,000\,\ms). The evolutionary parameter $\tau=\rm{age}/\tr$
(col.~12 of Table~\ref{tab5}), where $\tr$ is the relaxation time (\citealt{BinTre1987}), was found
to be a good tracer of dynamical states. In particular, significant flattening in cluster MF slopes 
due to dynamical effects such as mass segregation is expected to occur for $\tau\ga7$. Details on 
parameter correlation in the reference sample are given in \citet{DetAnalOCs}.

\begin{figure}
\resizebox{\hsize}{!}{\includegraphics{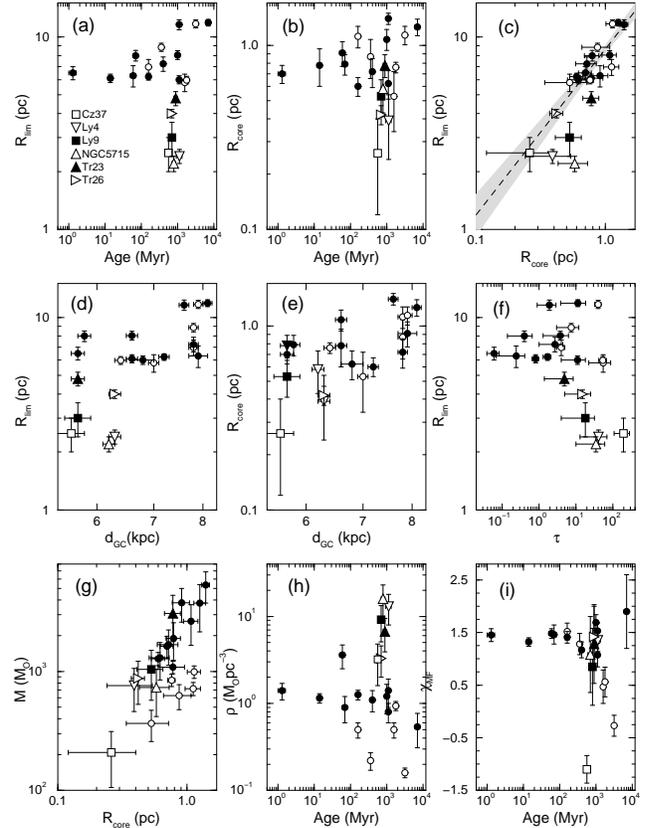}}
\caption[]{Relations involving structural and dynamical state parameters of OCs. Filled symbols:
OCs more massive than 1\,000\,\ms. Empty symbols: less massive OCs.}
\label{fig16}
\end{figure}

In panels (a) and (b) of Fig.~\ref{fig16} we compare core and limiting radii of the present OCs with
those of the reference sample in terms of cluster age. In all cases, both kinds of radii appear to be 
significantly smaller than those of nearby OCs of similar ages, especially in limiting radius. Tr\,23 
and Tr\,26 have $\rl\sim1/2$ of those in the reference sample, while for the remaining ones $\rl\sim1/3$. 
A similar effect occurs for Galactocentric distances (panels (d) and (e)). Core and limiting
radii in the reference sample are related by $\rl=(8.9\pm0.3)\times R_{\rm core}^{(1.0\pm0.1)}$
(panel (c)), which suggests that both kinds of radii undergo a similar scaling, at least for
$0.5\la\rc(pc)\la1.5$ and $5\la\rl(pc)\la15$. Except for Cz\,37 and Tr\,26, the remaining OCs do
not follow that relation. Considering that we may have underestimated \rl\ by 10--20\%
(Sect.~\ref{Simul}), this again suggests that the present OCs are either intrinsically small or have
been suffering important evaporation effects. The dependence of \rl\ on evolutionary parameter 
$\tau$ (panel (f)) supports this conclusion, since except for Tr\,23, the remaining OCs have 
$\tau$ significantly larger than 7 (\citealt{DetAnalOCs}). 

Relations presented in the bottom panels of Fig.~\ref{fig16} also indicate that we are dealing with 
OCs affected by tidal and low-contrast effects, especially in their outer parts. The relation of their 
(total) masses with core radii is consistent with that defined by nearby OCs (panel (g)), in both mass 
regimes. Mass density, on the other hand, that in the context of the present work is basically sensitive 
to limiting radius, seems to be exceedingly high with respect to nearby OCs. Except for Cz\,37 and Tr\,26, 
the remaining OCs have a mass density $\sim8$ times higher than that of nearby OCs. If mass values are 
reasonably well determined, as suggested by panel (g), a factor of $\sim2$ applied to their limiting radii 
would bring densities close to the typical values of nearby OCs. Interestingly, this correction would as
well put the deviant OCs back into the tight core and limiting radii relation (panel (c)). Even so,
they would still have core and limiting radii smaller than those of nearby OCs, with $\rc\la0.6$\,pc
and $\rl\la5.2$\,pc. Considering the discussion presented in Sect.~\ref{Simul}, most of the 
smaller limiting radii can be accounted for by dynamical effects, both internal
and external to the clusters.

Probably because of its innermost position in the Galaxy (Table~\ref{tab1}) and low mass
(Table~\ref{tab5}), Cz\,37 seems to be the OC most affected by tidal effects (and accelerated
dynamical evolution), since it shows the flattest MF slope (panel (i)), smallest core radius
and total mass in the present sample.

\section{Summary and conclusions}
\label{Conclu}

Low-contrast star clusters projected against central parts of the Galaxy present a
challenge to observational projects that intend to improve their census and derive
fundamental and dynamical parameters. High background levels 
may not allow detection of the external parts of clusters, because of low-surface brightness. 
In many cases, it may not be 
possible to detect faint clusters above the background at all (e.g. \citealt{DiskProp}).

Theoretical predictions indicate that poorly-populated OCs must by far be the majority of
the star cluster population in the Galaxy. Indeed, \citet{DiskProp} have found that the distribution
function for the number of OCs with $\eta$ observed member stars (considering all disc directions in the
Galaxy) follows $\phi_{OC}{(\eta)}=\frac{dN_{OC}}{d\eta}\propto e^{-\eta/95}$.
This implies that the fraction of Milky Way OCs with $\eta\la100$ observed member stars amounts
to about 65\%, while for $\eta\la300$ it increases to $\sim96\%$. Derivation of accurate fundamental and
structural parameters of poorly-populated OCs, especially those inside the Solar circle, is important
as well to test models of cluster tidal disruption.

In the present work a set of tools was applied - e.g. FS decontamination, CM filter and diagnostic-diagrams 
for structure and dynamical state - to six OCs projected not far 
from the Galactic center. The objects are NGC\,5715, Lyng\aa\,4, Lyng\aa\,9, Trumpler\,23, Trumpler\,26 
and Czernik\,37, whose near-IR CMDs are heavily contaminated by field stars, especially bulge ones. 
Probably because of this, these objects could not previously be studied in detail. Lyng\aa\,9, in particular, 
was considered to be a fluctuation of the dense field (\citealt{CJE05}). We work with wide-field 
extractions of 2MASS photometry. 

2MASS field-star decontaminated and colour-magnitude filtered photometry produced compelling evidence - 
in terms of CMDs with well-defined cluster sequences and stellar density profiles following King law -  
that the 6 objects are OCs about the Hyades age ($\rm 0.6\la age(Gyr)\la 1.3$) located $0.9 - 1.6$\,kpc 
inside the Solar Circle. In all cases core, and especially limiting radii, appear to be smaller than 
those of nearby OCs outside the Solar circle. We measured $\rm0.26\la\rc(pc)\la0.66$ and
$2.2\la\rl(pc)\la4.4$. 
 
Simulations of King-like OCs have shown that high background levels may have affected limiting radius 
estimates for the present OCs, in the sense that \rl\ can be underestimated by about 10--20\%. 
\rc, on the other hand, is almost insensitive to the range of background levels used in the simulations. 
Thus, background contamination alone cannot account for the small limiting radii measured in the present 
OCs, which are scaled down by factors of $1/3 - 1/2$, as compared to a sample of nearby OCs. Except for 
Cz\,37 which has a flat MF, the remaining OCs have Salpeter-like MFs, within uncertainties. Mass stored 
in observed MS and evolved stars amounts to $130\la M_{obs}(\ms)\la520$. Total masses, estimated by 
extrapolation of the MFs to 0.08\,\ms, are in the range $210\la M_{OC}(\ms)\la2700$. However we point 
out that total mass estimates may be upper limits because of dynamical evolution effects.

With the help of diagnostic diagrams that probe structure and dynamical state of OCs, we find evidence that 
the 6 objects have been suffering significant tidal effects (for ages older than $\sim600$\,Myr). Such  
tidal effects may have accelerated their dynamical evolution, especially in Czernik\,37, the innermost OC 
of the present sample with a flat MF. Large-scale mass segregation and low-mass 
star evaporation drive preferentially faint stars to large distances from the cluster centre. However, 
because dynamically-inflated cluster external regions end up having a low surface brightness, they tend 
to become indistinguishable from the background when projected against dense stellar fields. As an
observational consequence of these effects, poorly-populated OCs projected against the central parts of
the Galaxy tend to have smaller limiting radii than those measured in similar OCs located at high Galactic 
latitudes and/or far from the centre.

Arguments based only on observed CMD morphology may not be enough to establish the nature of objects 
in dense fields as star clusters. Instead, the present work shows that stronger constraints are provided 
by FS-decontaminated CMDs and the shape of CM-filtered RDPs, especially when RDPs of different magnitude 
domains present similar features. Besides, with the quantitative tools we have been developing in this 
series of works, it is becoming feasible to explore faint OCs in dense stellar fields with 2MASS photometry, 
provided care is taken to statistically identify probable member stars (to better define cluster stellar 
sequences on CMDs) and exclude stars with discordant colours (for more intrinsic cluster RDPs and MFs). 
Considering as well structure and dynamical state diagnostic-diagrams, it is becoming possible to disentangle 
high background levels from tidal effects in the outer parts of OCs.

\section*{acknowledgments}
We thank an anonymous referee for helpful suggestions.
This publication makes use of data products from the Two Micron All Sky Survey, which
is a joint project of the University of Massachusetts and the Infrared Processing and
Analysis Center/California Institute of Technology, funded by the National Aeronautics
and Space Administration and the National Science Foundation. This research has made 
use of the WEBDA database, operated at the Institute for Astronomy of the University 
of Vienna. We acknowledge support from the Brazilian Institution CNPq.

\label{lastpage}
\end{document}